%% file: outward.tex
\documentclass[sigconf]{acmart}

\usepackage{graphicx}
\usepackage{balance}  
\usepackage{booktabs}
\usepackage{amsmath}
\usepackage{amssymb}
\usepackage[linesnumbered,ruled,vlined]{algorithm2e}
\usepackage{balance}

\usepackage{graphicx}
\usepackage{subfig}
\usepackage{caption}
\usepackage{booktabs}
\usepackage{multirow}
\usepackage{epstopdf}
\usepackage{colortbl}
\usepackage{algpseudocode}
\usepackage{enumitem}
\usepackage{varwidth}
\usepackage{mathtools}
\DeclarePairedDelimiter\ceil{\lceil}{\rceil}
\DeclarePairedDelimiter\floor{\lfloor}{\rfloor}

\newtheorem{Lemma}{Lemma}
\newtheorem{Definition}{Definition}

\newcommand\I{\mathbb{I}}
\newcommand\E{\mathbb{E}}

\newcommand{\Var}{\mathrm{Var}}

\usepackage{xcolor}

\def\IE{\textsf{IE}}

\def\IICP{\textsf{IICP}}

\def\FPRAS{\textsf{FPRAS}}
\def\GSRA{\textsf{GSRA}}

\def\SiEIE{\textsf{RSA}}

\def\G{\mathcal{G}}

\def\Ext{\I_{out}}
\def\hExt{\hat \I_{out}}

\def\MC{\textsf{MC}}

\def\KDD{\textsf{INFEST}}

\def\MCb{\textsf{MC$_{\text{100K}}$}}
\def\MCe{\textsf{MC$_{\epsilon,\delta}$}}
\def\MCk{\textsf{MC$_{\text{10K}}$}}

\def\OInf{\textsf{SOIEA}}
\def\OBInf{\textsf{SIEA}}

\def\Ys{Y^{(S)}}
\def\pYs{Y'^{(S)}}
\def\Ms{M^{(S)}}
\def\Zs{Z^{(S)}}
\def\pZs{Z'^{(S)}}
\def\Ws{W^{(S)}}

\begin{document}
\title{Outward Influence and Cascade Size Estimation\\ in Billion-scale Networks}
\author{H. T. Nguyen, T. P. Nguyen}
\affiliation{%
	\institution{Virginia Commonwealth Univ.}
	\city{Richmond} 
	\state{VA} 
	\postcode{23220}
}
\email{{hungnt, trinpm}@vcu.edu}

\author{T. N. Vu}
\affiliation{%
	\institution{Univ. of Colorado, Boulder \& UC~Denver}
	\city{Boulder} 
	\state{CO} 
	\postcode{80309}
}
\email{tam.vu@colorado.edu }

\author{T. N. Dinh}
\orcid{0000-0003-4847-8356}
\affiliation{%
	\institution{Virginia Commonwealth Univ.}
	\city{Richmond} 
	\state{VA} 
	\postcode{23220}
}
\email{tndinh@vcu.edu}

\renewcommand{\shortauthors}{Nguyen et al.}

\begin{abstract}
Estimating cascade size and nodes' influence is a fundamental task in social, technological, and biological networks. Yet this task is extremely challenging due to the sheer size and the structural heterogeneity of networks. We investigate a new influence measure, termed \emph{outward influence} (OI), defined as the (expected) number of nodes that a subset of nodes $S$ will activate, \emph{excluding the nodes in $S$}. Thus, OI equals, the de facto standard measure, \emph{influence spread} of $S$  minus $|S|$. OI is not only more informative for nodes with small influence, but also, critical in designing new effective sampling and statistical estimation methods. 

Based on OI, we propose  SIEA/SOIEA, novel methods to estimate influence spread/outward influence \emph{at scale and with rigorous theoretical guarantees}.  The proposed methods are built on two novel components 1) \IICP{} an important sampling method for outward influence; and 2) RSA, a robust mean estimation method that minimize the number of samples through  analyzing variance and range of random variables. Compared to the state-of-the art for influence estimation, SIEA is  $\Omega(\log^4 n)$ times faster in theory and up to \emph{several orders of magnitude faster} in practice. For the first time, influence of nodes in the networks of billions of edges can be estimated with high accuracy within a few minutes. Our comprehensive experiments on real-world networks also give  evidence against the popular practice of using a fixed number, e.g. 10K or 20K, of samples to compute the ``ground truth'' for influence spread.
\end{abstract}

\maketitle

\newcommand*{\Scale}[2][4]{\scalebox{#1}{$#2$}}%
\newcommand*{\Resize}[2]{\resizebox{#1}{!}{$#2$}}%
\input{body/introduction}

\input{body/model}

\input{body/hardness}

\input{body/fpras}

\input{body/mean_est}

\input{body/infest}
\input{body/experiment}

\section{Related work}
\label{sec:work}
In a seminal paper \cite{Kempe03}, Kempe et al. formulated and generalized two important influence diffusion models, i.e. Independent Cascade (IC) and Linear Threshold (LT). This work has motivated a large number of follow-up researches on information diffusion \cite{Kempe05,Chen10,Borgs14,Lucier15,Cohen14,Ohsaka16} and applications in multiple disciplines \cite{Leskovec07,Ioannides04,Krause08}. Kempe et al. \cite{Kempe03} proved the monotonicity and submodularity properties of influence as a function of sets of nodes. Later, Chen et al. \cite{Chen10} proved that computing influence under these diffusion models is \#P-hard.

Most existing works uses the naive influence cascade simulations to estimate influences \cite{Kempe03,Leskovec07,Chen10,Lucier15}. Most recently, Lucier et al. \cite{Lucier15} proposed an estimation algorithm with rigorous quality guarantee for a single seed set. The main idea is guessing a small interval of size $(1+\epsilon)$ that the true influence falls in and verifying whether the guess is right with high probability. However, their approach is not scalable due to a main drawback that the guessed intervals are very small, thus, the number of guesses as well as verifications made is huge. As a result, the method in \cite{Lucier15} can only run for small dataset and still takes hours to estimate a single seed set. They also developed a distributed version on MapReduce however, graph algorithms on MapReduce have various serious issues \cite{Gupta13,Lin10}.

Influence estimation oracles are developed in \cite{Cohen14,Ohsaka16} which take advantage of sketching the influence to preprocess the graph for fast queries. Cohen et al. \cite{Cohen14} use the novel bottom-$k$ min-hash sketch to build combined reachability sketches while Ohsaka et al. in \cite{Ohsaka16} adopt the reverse influence sketches. \cite{Ohsaka16} also introduces the reachability-true-based technique to deal with dynamic changes in the graphs. However, these methods require days for preprocessing in order to achieve fast responses for multiple queries.

There have also been increasing interests in many related problems. \cite{Cha09,Goyal10} focus on designing data mining or machine learning algorithms to extract influence cascade model parameters from real datasets, e.g. action logs. Influence Maximization, which finds a seed set of certain size with the maximum influence among those in the same size, found many real-world applications and has attracted a lot of research work \cite{Kempe03,Leskovec07,Chen10,Nguyen13icdm, Borgs14,Tang15,Nguyen162,Nguyen163}.

\input{body/conclusion}

\bibliographystyle{ACM-Reference-Format}
\bibliography{social,pids,targetedIM,budgetedIM,infEst}

\appendix
\subsection*{Proof of Lemma~\ref{lem:submod}}	
%
%
Recall that on a sampled graph $g \sim \G$, for a set $S \subseteq V$, we denote $r^{(o)}_g(S)$ to be the set of nodes, excluding the ones in $S$, that are reachable from $S$ through live edges in $g$, i.e. $r^{(o)}_g(S) = r_g(S)\backslash S$. Alternatively, $r^{(o)}_g(S)$ is called the outward influence cascade of $S$ on sample graph $g$ and, consequently, we have,
\begin{align}
	\Ext(S) = \sum_{g \sim \G} |r^{(o)}_g(S)| \Pr[g \sim \G].
	\label{eq:compute_ie}
\end{align}

It is sufficient to show that $|r^{(o)}_g(S)|$ is submodular, as $\Ext(S)$ is a linear combination of submodular functions. Consider a sample graph $g \sim \G$, two sets $S, T$ such that $S \subseteq T \subseteq V$ and $v \in V \backslash T$. We have three possible cases:
\begin{itemize}
	\item \textbf{Case $v \in r^{(o)}_g(S)$}: then $v \in r^{(o)}_g(T)$ since $S \subseteq T$ and $v \notin T$. Thus, we have the following,
	\begin{align}
		& r^{(o)}_g(S \cup \{v\}) - r^{(o)}_g(S) \nonumber \\
		& \qquad\qquad= r^{(o)}_g(T \cup \{v\}) - r^{(o)}_g(T) = -1.
	\end{align}
	\item \textbf{Case $v \notin r^{(o)}_g(S)$ but $v \in r^{(o)}_g(T)$}: We have that,
	\begin{align}
		& r^{(o)}_g(S \cup \{v\}) - r^{(o)}_g(S) \nonumber \\
		& \qquad\qquad = |r^{(o)}_g(\{v\}) \backslash (r^{(o)}_g(S) \cup S)| \geq 0,
	\end{align}
	while $r^{(o)}_g(T \cup \{v\}) - r^{(o)}_g(T) = -1$. Thus,
	\begin{align}
		& r^{(o)}_g(S \cup \{v\}) - r^{(o)}_g(S) \nonumber \\
		& \qquad \qquad > r^{(o)}_g(T \cup \{v\}) - r^{(o)}_g(T).
	\end{align}
	\item \textbf{Case $v \notin r^{(o)}_g(T)$}: Since $\forall u \in r^{(o)}_g(S) \cup S$, we have either $u \in r^{(o)}_g(T)$ or $u \in T$ or $r^{(o)}_g(S) \cup S \subseteq r^{(o)}_g(T) \cup T$, and thus,
	\begin{align}
		& r^{(o)}_g(S \cup \{v\}) - r^{(o)}_g(S) \nonumber \\
		& \qquad\qquad= |r^{(o)}_g(\{v\}) \backslash (r^{(o)}_g(S)\cup S)| \nonumber \\
		& \qquad\qquad\geq |r^{(o)}_g(\{v\}) \backslash (r^{(o)}_g(T) \cup T)| \nonumber \\
		& \qquad\qquad= r^{(o)}_g(T \cup \{v\}) - r^{(o)}_g(T).
	\end{align}
\end{itemize}
In all three cases, we have,
\begin{align}
	\label{eq:sub_g}
	& r^{(o)}_g(S \cup \{v\}) - r^{(o)}_g(S) \nonumber \\
	& \qquad\geq r^{(o)}_g(T \cup \{v\}) - r^{(o)}_g(T).
\end{align}
Applying Eq.~\ref{eq:sub_g} on all possible $g \sim \G$ and taking the sum over all of these inequalities give
\vspace{-0.05in}
\begin{align}
	\sum_{g \sim \G}(r^{(o)}_g&(S \cup \{v\}) - r^{(o)}_g(S)) \Pr[g \sim \G] \nonumber \\
	&\geq \sum_{g \sim \G}(r^{(o)}_g(T \cup \{v\}) - r^{(o)}_g(T)) \Pr[g \sim \G], \nonumber
\end{align}
\vspace{-0.2in}

or,
\vspace{-0.05in}
\begin{align}
	\Ext(S \cup \{v\}) - \Ext(S) \geq \Ext(T \cup \{v\}) - \Ext(T).
\end{align}
That completes the proof.

\subsection*{Proof of Lemma~\ref{lem:sim_prob}}
Let $\Omega^+_{W}$ be the probability space of all possible cascades from $S$. For any cascade $\Ws{} \supseteq S$, the probability of that cascade in $\Omega^+_{W}$ is given by
\vspace{-0.05in}
\[
\Pr[\Ws{} \in \Omega^+_W] = \sum_{g \in \Omega_\G, g \leadsto \Ws{}} \Pr[g \in \Omega_\G],
\]
\vspace{-0.1in}

\noindent where $g \leadsto \Ws{}$ means that $\Ws{}$ is the set of reachable nodes from $S$ in $g$.

Let $\Omega_W$ be the probability space of non-trivial cascades. According to the Stage 1 in \IICP{}, the probability of the trivial cascade is:
\vspace{-0.05in}
\[
\Pr[S \in \Omega_W] = 0.
\]
Comparing to the mass of cascades in $\Omega^+_{W}$, the probability mass of the trivial cascade $S$ in $\Omega_W$ is redistributed proportionally to other cascades in $\Omega_W$. Specifically,  according to line 2 in \IICP{}, the probability mass of all the non-trivial cascades in $\Omega_{W}$ is multiplied by a factor $1/\beta_0$. Thus,
\vspace{-0.05in}
\[	
\Pr[\Ws{} \in \Omega^+_W] = \Pr[\Ws{} \in \Omega_W] \cdot \beta_0\ \ \forall \Ws{} \neq S.
\]	
It follows that
\vspace{-0.05in}
\begin{align}
	\Ext(S) & = \sum_{\Ws{} \in \Omega^+_W} |\Ws{} \setminus S| \cdot \Pr[\Ws{} \in \Omega^+_W] \\
	& = \sum_{\Ws{} \in \Omega_W} |\Ws{} \setminus S|
	\cdot \Pr[\Ws{} \in \Omega^W]
	\beta_0\\
	& = \E[|\Ws{}|] \cdot \beta_0 = \E[\Ys{}] \cdot \beta_0.
\end{align}
\vspace{-0.1in}

\noindent We note that for $\Ws{} = S$, $|\Ws{} \setminus S| = 0$. Thus the difference in the probability masses between the two probabilistic spaces does not affect the 2nd step.

\subsection*{Proof of Theorem~\ref{theo:stop}}
We will equivalently prove two probabilistic inequalities:
\begin{align}
	\label{eq:ineq_1}
	\Pr[\hat \mu_X < (1-\epsilon)\mu_X] \leq \frac{\delta}{2},
\end{align}
and
\begin{align}
	\label{eq:ineq_2}
	\Pr[\hat \mu_X > (1+\epsilon)\mu_X] \leq \frac{\delta}{2}.
\end{align}
\indent \textit{Prove Eq.~\ref{eq:ineq_1}.}
We first realize that at termination point of Alg.~\ref{alg:stop}, due to the stopping condition $h = \sum_{j = 1}^{T} X_j \geq \Upsilon$ and $X_j \leq b, \forall j$, the following inequalities hold,
\begin{align}
	\label{eq:eq34}
	\Upsilon \leq \sum_{j = 1}^T X_j \leq \Upsilon + b.
\end{align}
The left hand side of Eq.~\ref{eq:ineq_1} is rewritten as follows,
\begin{align}
	\Pr[\hat \mu_X < (1-\epsilon)\mu_X] & = \Pr\Big[\frac{\sum_{j = 1}^{T} X_j}{T} < (1-\epsilon)\mu_X\Big] \\
	& = \Pr\Big[\sum_{j = 1}^{T} X_j < (1-\epsilon)\mu_X T\Big] \\
	& \leq \Pr[\Upsilon < (1-\epsilon)\mu_X T].
	\label{eq:eq37}
\end{align}
The last inequality is due to our realization in Eq.~\ref{eq:eq34}. Assume that $\epsilon < 1$ and $\mu_X > 0$, let denote $L_1 = \ceil{\frac{\Upsilon}{(1-\epsilon)\mu_X}}$. We then have,
\begin{align}
	\label{eq:eq38}
	L_1 \geq \frac{\Upsilon}{(1-\epsilon)\mu_X} \Rightarrow \frac{\Upsilon}{L_1} \leq (1-\epsilon)\mu_X,
\end{align}
and
\begin{align}
	\label{eq:eq39}
	L_1 > \frac{\Upsilon}{\mu_X} > (2+\frac{2}{3}\epsilon) \ln (\frac{2}{\delta}) \frac{1}{\epsilon'^2\mu_X } (b-a).
\end{align}
Thus, from Eq.~\ref{eq:eq37}, we obtain,
\begin{align}
	\Pr[\hat \mu_X < (1-\epsilon)\mu_X] & \leq \Pr[L_1 \leq T] = \Pr\Big[\sum_{j = 1}^{L_1}X_j \leq \sum_{j = 1}^{T} X_j\Big] \nonumber \\
	& \leq \Pr\Big[\sum_{j = 1}^{L_1}X_j \leq \Upsilon + b\Big]
	\label{eq:eq40}\\
	& \leq \Pr\Big[\frac{\sum_{j = 1}^{L_1}X_j}{L_1} \leq \frac{\Upsilon + b}{L_1}\Big],
	\label{eq:eq41}
\end{align}
where the second inequality is due to Eq.~\ref{eq:eq34}. Note that $\frac{\sum_{j = 1}^{L_1}X_j}{L_1}$ is an estimate of $\mu_X$ using the first $L_1$ random variables $X_1, \dots, X_{L_1}$. Furthermore, from Eq.~\ref{eq:eq38} that $\frac{\Upsilon}{L_1} \leq (1-\epsilon)\mu_X$, we have,
\begin{align}
	\frac{\Upsilon+b}{L_1} \leq (1-\epsilon)\mu_X + \frac{b}{L_1} = (1 - \epsilon + \frac{b}{L_1 \mu_X})\mu_X.
\end{align}
Since $L_1 > (2+\frac{2}{3}\epsilon) \ln (\frac{2}{\delta}) \frac{1}{\epsilon'^2\mu_X } (b-a)$ from Eq.~\ref{eq:eq39},
\begin{align}
	\frac{\Upsilon+b}{L_1} \leq \Big (1 - \epsilon + \frac{\epsilon^2 b}{(2+\frac{2}{3}\epsilon) \ln (\frac{2}{\delta})(b-a)} \Big )\mu_X = (1-\epsilon') \mu_X.
\end{align}
Plugging these into Eq.~\ref{eq:eq41}, we obtain,
\begin{align}
	\Pr[\hat \mu_X < (1-\epsilon)\mu_X] & \leq \Pr\Big[\sum_{j = 1}^{L_1} X_j \leq (1-\epsilon')\mu_X L_1\Big].
\end{align}
Now, apply the Chernoff-like bound in Eq.~\ref{eq:gen_bound_1} with $T = L_1$ and note that $L_1 > (2+\frac{2}{3}\epsilon) \ln (\frac{2}{\delta}) \frac{1}{\epsilon'^2\mu_X } (b-a) > 2 \ln (\frac{2}{\delta}) \frac{1}{\epsilon'^2\mu_X } (b-a)$, we achieve,
\begin{align}
	\Pr[\hat \mu_X < (1-\epsilon)\mu_X] & \leq \exp \Big ( - \frac{\epsilon'^2 L_1 \mu_X}{2(b-a)}\Big) \\
	& \leq \exp \Big( - \frac{\epsilon'^2 2 \ln (\frac{2}{\delta})\frac{1}{\epsilon'^2 \mu_X} (b-a)}{2 (b-a)} \Big) \nonumber \\
	& = \frac{\delta}{2}.
\end{align}
That completes the proof of Eq.~\ref{eq:ineq_1}.

\indent \textit{Prove Eq.~\ref{eq:ineq_2}.}
The left hand side of Eq.~\ref{eq:ineq_2} is rewritten as follows,
\begin{align}
	\Pr[\hat \mu_X > (1+\epsilon)\mu_X] & = \Pr\Big[\sum_{j = 1}^{T} X_j > (1+\epsilon)\mu_X T\Big] \\
	& \leq \Pr[\Upsilon + b > (1+\epsilon)\mu_X T],
	\label{eq:eq48}
\end{align}
where the last inequality is due to our observation that $\sum_{j = 1}^{T} X_j \leq \Upsilon + b$. Under the same assumption that $0 < \mu_X \leq \frac{b}{1+\epsilon}$, we denote $L_2 = \floor{\frac{\Upsilon+b}{(1+\epsilon)\mu_X}}$. We then have,
\begin{align}
	\label{eq:eq49}
	L_2 \geq \frac{\Upsilon}{(1+\epsilon)\mu_X} = (2+\frac{2}{3}\epsilon) \ln (\frac{2}{\delta}) \frac{1}{\epsilon'^2\mu_X } (b-a),
\end{align}
and
\begin{align}
	& L_2 \leq \frac{\Upsilon+b}{(1+\epsilon)\mu_X} \Rightarrow \frac{\Upsilon + b}{L_2} \geq (1+\epsilon)\mu_X \\
	\Rightarrow \text{ } & \frac{\Upsilon}{L_2} \geq (1+\epsilon)\mu_X - \frac{b}{L_2} = (1+\epsilon - \frac{b}{L_2 \mu_X})\mu_X \\
	\Rightarrow \text{ } & \frac{\Upsilon}{L_2} \geq \Big(1+\epsilon - \frac{\epsilon^2 b}{(2+\frac{2}{3}\epsilon)\ln (\frac{2}{\delta})(b-a)}\Big)\mu_X = (1+\epsilon')\mu_X
	\label{eq:eq52}
\end{align}
Thus, from Eq.~\ref{eq:eq48}, we obtain,
\begin{align}
	\Pr[\hat \mu_X > & (1+\epsilon)\mu_X] \leq \Pr[L_2 \geq T] = \Pr\Big[\sum_{j = 1}^{L_2} X_j \geq \sum_{j = 1}^{T} X_j\Big] \nonumber \\
	& \leq \Pr\Big[\sum_{j = 1}^{L_2} X_j \geq \Upsilon\Big] = \Pr\Big[\frac{\sum_{j = 1}^{L_2} X_j}{L_2} \geq \frac{\Upsilon}{L_2}\Big] \\
	& \leq \Pr\Big[\frac{\sum_{j = 1}^{L_2} X_j}{L_2} \geq (1+\epsilon')\mu_X\Big]
\end{align}
where the last inequality follows from Eq.~\ref{eq:eq52}. By applying another Chenoff-like bound from Eq.~\ref{eq:gen_bound_2} combined with the lower bound on $L_2$ in Eq.~\ref{eq:eq49}, we achieve,
\begin{align}
	\Pr[\hat \mu_X > (1+\epsilon)\mu_X] \leq \exp \big(-\frac{\epsilon'^2 L_2 \mu_X}{(2+\frac{2}{3}\epsilon)(b-a)} \big) = \frac{\delta}{2},
\end{align}
which completes the proof of Eq.~\ref{eq:ineq_2}.

Follow the same procedure as in the proof of Eq.~\ref{eq:ineq_2}, we obtain the second statement in the theorem that,
\begin{align}
	\Pr[T \leq (1+\epsilon)\Upsilon/\mu_X] > 1 - \delta/2,
\end{align}
which completes the proof of the whole theorem.

\textbf{More elaboration on the hold in \cite{Dagum00}.} The stopping rule algorithm in \cite{Dagum00} is described in Alg.~\ref{alg:stop_old}.
\begin{algorithm} \small
	\caption{Stopping Rule Algorithm \cite{Dagum00}}
	\label{alg:stop_old}
	\KwIn{Random variables $X_1, X_2, \dots$ and $0 < \epsilon,\delta < 1$}
	\KwOut{An $(\epsilon,\delta)$-approximate of $\mu_X = E[X_i] $}
	Compute: $\Upsilon_1 = 1+ (1+\epsilon) 4 (e-2) \ln (\frac{2}{\delta}) \frac{1}{\epsilon^2}$;\\
	Initialize $h = 0, T = 0$;\\
	\While{$h < \Upsilon_1$}{
		$ h \leftarrow h + X_T, T \leftarrow T+1$;\\
	}
	\textbf{return} $\hat \mu_X = \Upsilon_1/T$;
\end{algorithm}

The algorithm first computes $\Upsilon_1$ and then, generates samples $X_j$ until the sum of their outcomes exceed $\Upsilon_1$. Afterwards, it returns $\Upsilon_1/T$ as the estimate. Apparently, $\Upsilon_1/T$ is a biased estimate of $\mu_X$ since $\sum_{j=1}^T X_j \geq \Upsilon_1$.

An important realization for this algorithm from our proof of Theorem~\ref{theo:stop} is that $\Upsilon_1 \leq \sum_{j = 1}^{T} X_j \leq \Upsilon_1 + b$ with $b = 1$ for $[0,1]$ random variables. In section~5 of \cite{Dagum00}, following the proof of $\Pr[\hat \mu_X > (1+\epsilon)\mu_X] \leq \delta/2$ to prove $\Pr[\hat \mu_X < (1-\epsilon)\mu_X] \leq \delta/2$, there is step that derives as follows: $\Pr[L_1 \leq T] = \Pr \left [\sum_{j=1}^{L_1} X_j \leq \sum_{j=1}^{T} X_j \right ] = \Pr \left [\sum_{j=1}^{L_1} X_j \leq \Upsilon_1 \right]$ where $L_1$ is a predefined number, i.e. $L_1 = \floor{\frac{\Upsilon_1}{(1-\epsilon)\mu_X}}$. However, since $\Upsilon_1 \leq \sum_{j = 1}^{T} X_j \leq \Upsilon_1 + b$, the last equality does not hold. This is based on Eq.~\ref{eq:eq40} with the correct expression being $\Pr\Big[\sum_{j = 1}^{L_1}X_j \leq \Upsilon + b\Big]$ instead of $\Pr\Big[\sum_{j = 1}^{L_1}X_j \leq \Upsilon\Big]$.


\subsection*{Proof of Theorem~\ref{th:extinf}}

\indent \textsf{[Proof of Part (1)]} If $\epsilon \geq 1/4$, then \SiEIE{} only runs \GSRA{} and hence, from Theorem~\ref{theo:stop}, the returned solution satisfies the precision requirement. Otherwise, since the first steps is literally applying \GSRA{} with $\sqrt{\epsilon} < 1/2,\delta/3$, we have,
\begin{align}
	\Pr[\mu_X(1-\sqrt{\epsilon}) \leq \hat \mu'_X \leq \mu_X(1+\sqrt{\epsilon})] \geq 1-\delta/3
\end{align}
We prove that in step 2, $\hat \rho_X \geq \rho_X/2$. Let define the random variables $\xi_i = (X'_{2i-1} - X'_{2i})^2/2, i = 1, 2, \dots$ and thus, $\E[\xi_i] = \Var[X]$. Consider the following two cases.

\begin{enumerate}	
	\item  \textbf{If $\Var[X] \geq \epsilon \mu_X (b-a)$}, consider two sub-cases:
	\begin{enumerate}
		\item If $\Var[X] \geq 2(1-\sqrt{\epsilon})\epsilon \mu_X (b-a)$, then since $N_\sigma = \Upsilon_2\epsilon/\hat \mu'_X \geq \frac{2}{1-\sqrt{\epsilon}}(1+\ln(\frac{3}{2})/\ln(\frac{2}{\delta}))\Upsilon\epsilon/\mu_X$, applying the Chernoff-like bound in Eq.~\ref{eq:eq33} gives,
		\begin{align}
			\Pr[\Var[X]/2 \leq \Delta/N_\sigma] \geq 1 - \delta/3
		\end{align}
		Thus, $\hat \rho_X \geq \Var[X]/2 = \rho_X/2$ with a probability of at least $1-\delta/3$.
		\item If $\Var[X] \leq 2(1-\sqrt{\epsilon})\epsilon\mu_X (b-a)$, then $\epsilon \mu_X(b-a) \geq \Var[X]/(2(1-\sqrt{\epsilon}))$ and therefore, $\hat \rho_X \geq \epsilon \hat \mu'_X (b-a) \geq (1-\sqrt{\epsilon})\epsilon \mu_X(b-a) \geq \Var_X/2 = \rho_X/2$.
	\end{enumerate}
	
	\item \textbf{If $\Var[X] \leq \epsilon \mu_X (b-a)$}, it follows that $\hat \rho_X \geq \epsilon \hat \mu_X \geq \rho_X (1-\min\{\sqrt{\epsilon},1/2\})$ with probability at least $1-\delta/3$.
	
\end{enumerate}
Thus, after steps 1 and 2, $2\frac{1+\sqrt{\epsilon}}{1-\sqrt{\epsilon}}\hat \rho_X/{\hat\mu}'^2_X \geq \rho_X/\mu^2_Z$ with probability at least $1-\delta/3$. In step 3, since $T = \Upsilon_2\hat \rho_X/({\hat\mu}'^2_X (b-a)) \geq (1+\ln(\frac{3}{2})/\ln(\frac{2}{\delta}))\Upsilon\rho_X/(\mu^2_X(b-a))$ and hence, applying the Chernoff-like bound in Eq.~\ref{eq:eq35} again gives,
\begin{align}
	\Pr[\mu_X(1-\epsilon) \leq \hat \mu_X \leq \mu_X(1+\epsilon)] \geq 1 - 2\delta/3.
\end{align}
Accumulating the probabilities, we finally obtain,
\begin{align}
	\Pr[\mu_X(1-\epsilon) \leq \hat\mu_X \leq \mu_X(1+\epsilon)] \geq 1 - \delta,
\end{align}
This completes the proof of part (1).

\vspace{0.1in}
\noindent \textsf{[Proof of Part (2)]}
The \SiEIE{} algorithm may fail to terminate after using $\mathcal{O}(\Upsilon \rho_X/(\mu^2_X(b-a)))$ samples if either: 
\begin{enumerate}
	\item The \GSRA{} algorithm fails to return an $(\sqrt{\epsilon},\delta/3)$-approximate $\hat \mu'_X$ with probability at most $\delta/2$, or,
	\item In step 2, for $\Var[X] \leq 2(1-\sqrt{\epsilon})\epsilon \mu_X (b-a)$, $\hat{\rho}_X$ is not $\mathcal{O}(\epsilon \mu_X (b-a))$ with probability at most $\delta/2$.				
\end{enumerate}

From Theorem~\ref{theo:stop}, with $T = (1+\epsilon)\Upsilon/\mu_X = \mathcal{O}(\Upsilon \rho_X/(\mu^2_X(b-a)))$, the first case happens with probability at most $\delta/2$. In addition, we can show similarly to Theorem~\ref{theo:stop} that if $\Var[X] \leq 2 \epsilon \mu_X (b-a)$, then,
\begin{align}
	\Pr[\Delta/T \geq 4\epsilon \mu_X (b-a)] \leq \exp(-T\epsilon \mu_X (b-a)/2).
\end{align}
Thus, for $T \geq 2 \Upsilon \epsilon/\mu_X$, we have $\Pr[\Delta/T \geq 4\epsilon \mu_X] \leq \delta/2$.

\subsection*{Proof of Lemma~\ref{lem:var}}

We start with the computation of $\Var[Z^{(S)}]$ with a note that $\E[Z^{(S)}] = \I(S)$,
\begin{align}
	&\Var[Z^{(S)}] = \sum_{z= \beta_0 + |S|}^{(n-|S|)\beta_0 + |S|} (z - \E[Z^{(S)}])^2 \Pr[Z^{(S)} = z] \nonumber \\
	& = \sum_{y = 1}^{n-|S|} (y \beta_0 + |S| - \I(S))^2 \Pr[Y^{(S)} = y] \nonumber \\
	& = \sum_{y = 1}^{n-|S|} (y \beta_0 - \Ext(S)\beta_0 + \Ext(S) \beta_0 + |S| - \I(S))^2 \Pr[Y^{(S)} = y] \nonumber \\
	& = \beta_0^2 \sum_{y = 1}^{n-|S|} (y-\Ext(S))^2\Pr[Y^{(S)} = y] \nonumber \\
	& \text{ \ } + \sum_{y = 1}^{n-|S|}(\Ext(S)\beta_0 + |S| - \I(S))^2 \Pr[Y^{(S)} = y] \nonumber \\
	& \text{ \ }+ 2\beta_0 \sum_{y = 1}^{n-|S|} (y-\Ext(S))(\Ext(S)\beta_0 + |S| - \I(S)) \Pr[Y^{(S)} = y] \nonumber
\end{align}

Since $Y^{(S)} \geq 1$ and $\Pr[Y^{(S)} = y] = \frac{\Pr[M^{(S)} = y+|S|]}{\beta_0}$, we have,
\begin{align}
	&\sum_{y = 1}^{n-|S|}(y - \Ext(S))^2\Pr[Y^{(S)} = y] \nonumber \\
	& \qquad = \frac{1}{\beta_0}\sum_{m = 1+|S|}^{n}(m - \E[M^{(S)}])^2\Pr[M^{(S)} = m] \nonumber \\
	& \qquad = \frac{1}{\beta_0}\sum_{m = |S|}^{n}(m - \E[M^{(S)}])^2\Pr[M^{(S)} = m] - \frac{1}{\beta_0}\Ext^2(S)(1-\beta_0) \nonumber \\
	& \qquad = \frac{1}{\beta_0} (\Var[M^{(S)}] - \Ext^2(S)(1-\beta_0)),
\end{align}
and,

\balance		
\begin{align}
	&\sum_{y = 1}^{n-|S|} \beta_0 (y-\Ext(S))(\Ext(S)\beta_0 + |S| - \I(S)) \Pr[Y^{(S)} = y] \nonumber \\
	&= (\Ext(S)\beta_0 + |S| - \I(S)) \sum_{y = 1}^{n-|S|} (y-\Ext(S)) \Pr[Y^{(S)} = y] \nonumber \\ 
	&= (\Ext(S)\beta_0 + |S| - \I(S)) \Ext(S)(1/\beta_0 - 1).
\end{align}

Plug these back in the $\Var[Z^{(S)}]$, we obtain,
\begin{align}
	\Var[Z^{(S)}] & = \beta_0 (\Var[M^{(S)}] - \Ext^2(S)(1-\beta_0)) \nonumber \\
	& \qquad + (\Ext(S)\beta_0 + |S| - \I(S))^2 \nonumber \\
	& \qquad + 2\beta_0(\Ext(S)\beta_0 + |S| - \I(S)) \Ext(S)(1/\beta_0 - 1) \nonumber \\
	& = \beta_0 \cdot \Var[M^{(S)}] - (1-\beta_0)\Ext^2(S)
\end{align}
That completes the computation.

\end{document}

%% file: body/introduction.tex
\vspace{-0.1in}
\section{Introduction}
\label{sec:intro}

In the past decade, a massive amount of data on human interactions has shed light on various cascading processes from the propagation of information and influence \cite{Kempe03} to the outbreak of diseases \cite{Leskovec07}. These cascading processes can be modeled in graph theory through the abstraction of the network as a graph $G=(V, E)$ and a \emph{diffusion model} that describes how the cascade proceeds into the network from a prescribed subset of nodes. A fundamental task in analyzing those cascades is to estimate the cascade size,  also known as \emph{influence spread} in social networks.  This task is the foundation of the solutions for many applications  including  viral marketing \cite{Kempe03, Tang14, Tang15,  Nguyen163}, estimating users' influence \cite{Du13,Lucier15}, optimal vaccine allocation \cite{Preciado13}, identifying critical nodes in the network \cite{Dinh15infocom},  and many others. 
Yet this task becomes computationally challenging in the face of the nowadays social networks that may consist of billions of nodes and edges.

Most of the existing work in network cascades uses stochastic diffusion models and estimates the influence spread through sampling \cite{Kempe03,Cohen14,Dinh15infocom,Tang15, Lucier15, Ohsaka16}. The common practice is to use a fixed number of samples, e.g. 10K or 20K \cite{Kempe03, Tang15, Cohen14, Ohsaka16}, to estimate the expected size of the cascade, aka \emph{influence spread}. Not only is there no single sample size that works well for all networks of different sizes and topologies, but those approaches also do not provide any accuracy guarantees. Recently, Lucier et al. \cite{Lucier15} introduced \KDD{}, the first estimation method that comes with accuracy guarantees. Unfortunately, our experiments suggest that \KDD{} does not perform well in practice, taking hours on networks with only few thousand nodes.  \emph{Will there be a rigorous method to estimate the  cascade size in billion-scale networks?}

\setlength\tabcolsep{3pt}
\def\arraystretch{1}
\begin{figure}[t]
	\begin{minipage}[t]{0.3\textwidth} \centering
		\vspace{-0.75in}
		\begin{tabular}{l|r|r}
			\toprule
			$S$ & Influence $\I(S)$ & Outward Inf. $\Ext(S)$ \\
			\midrule
			$\{u\}$ & $1+p+2p^2= 1.12$ & $p+2p^2=0.12$ \\
			$\{v\}$ & $1+2p=1.20$ & $2p=0.20$ \\
			$\{w\}$ & $     1.00$ & $0.00$ \\
			\bottomrule
		\end{tabular}
	\end{minipage}%
	\begin{minipage}[t]{0.25\textwidth} \centering
		\includegraphics[width=0.8in]{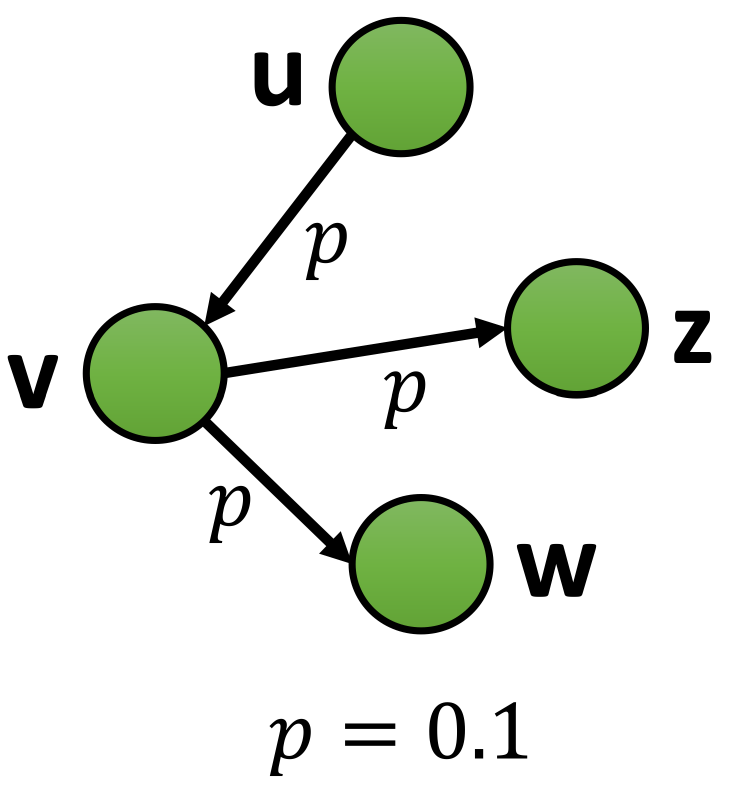}
		\vspace{-5in}
	\end{minipage}
	\vspace{-0.2in}
	\caption{\small \underline{Left}: the influence of nodes under IC model. The influence of all nodes are roughly the same, despite that $w$ is much less influential than $u$ and $v$. \underline{Right}: Outward influence is better at reflecting the relative influence of the nodes. $w$ has the least outward influence, $0$, while $v$'s is nearly twice as that of $u$}
	\label{fig:example}
	\vspace{-0.2in}
\end{figure}

In this paper, we investigate efficient estimation methods for nodes' influence under stochastic cascade models \cite{Daley01, Kempe03, Du13}.  First, we introduce a new influence measure, called \emph{outward influence} and defined as $\I_{out}(S)=\I(S) - |S|$, where $\I(S)$ denotes the influence spread. The new measure excludes the self-influence artifact in influence spread, making it \emph{more effective in comparing relative influence of nodes}. As shown in Fig.~\ref{fig:example}, the influence spread of the nodes are roughly the same, $1$. In contrast, the outward influence of nodes $u, v$ and $w$ are $0.12, 0.20$, and $0.00$, respectively. Those values correctly reflect the intuition that $w$ is the least influential nodes and $v$ is nearly twice as influential as $u$. 

%

More importantly,  the outward influence measure inspires novel methods,  termed SIEA/SOIEA, to estimate influence spread/outward influence \emph{at scale and with rigorous theoretical guarantees}. Both \OInf{} and \OBInf{} guarantee
\emph{arbitrary small relative error with  high probability} within an $O(n)$ observed influence.  The proposed methods are built on two novel components 1) \IICP{} an important sampling method for outward influence; and 2) RSA, a robust mean estimation method that minimize the number of samples through  analyzing variance and range of random variables.  \IICP{} focuses only on \emph{non-trivial cascades} in which at least one node outside the seed set must be activated. As each \IICP{} generates cascades of size at least two and outward influence of at least one, it leads to smaller variance and much faster convergence to the mean value. Under the well-known independent cascade model \cite{Kempe03}, $\OInf{}$ is \emph{$\Omega(\log^4 n)$ times faster} than  the state-of-the-art \KDD{}  \cite{Lucier15} in theory and is \emph{four to five orders of magnitude faster} than both \KDD{} and the naive Monte-Carlo sampling. For other stochastic models, such as continuous-time diffusion model \cite{Du13}, LT model \cite{Kempe03}, SI, SIR, and variations \cite{Daley01}, RSA can be applied directly to estimate the influence spread, given a Monte-Carlo sampling procedure, or, better, with an extension of \IICP{} to the model.

Our contributions are summarized as follows.
\begin{itemize}
	\item We introduce a new influence measure, called \textit{Outward Influence} which is more effective in differentiating nodes' influence. We investigate the characteristics of this new measure including non-monotonicity, submodularity, and \#P-hardness of computation.
 \item Two fully polynomial time randomized approximation schemes (\FPRAS) \OBInf{}  and \OInf{} to provide $(\epsilon, \delta)$-approximate for influence spread and outward influence with only an $O(n)$ observed influence in total. Particularly, $\OInf{}$, our algorithm to estimate influence spread, is \emph{$\Omega(\log^4 n)$ times faster} than  the state-of-the-art \KDD{}  \cite{Lucier15} in theory and is \emph{four to five orders of magnitude faster} than both \KDD{} and the naive Monte-Carlo sampling.
\item The robust mean estimation algorithm, termed \SiEIE{}, a building block of \OBInf{},  can be used to estimate influence spread under \emph{other stochastic diffusion models}, or, in general, mean of bounded random variables of unknown distribution. \SiEIE{} will be our favorite statistical algorithm moving forwards. 
	\item We perform comprehensive experiments on both real-world and synthesis networks with size up to 65 million nodes and \emph{1.8 billion edges}. Our experiments indicate the superior of our algorithms in terms of both accuracy and running time in comparison to  the naive Monte-Carlo and the state-of-the-art methods. The results also give \emph{evidence against the practice of using a fixed number of samples} to estimate the cascade size. For example, using 10000 samples to estimate the influence will deviate up to 240\% from the ground truth in a Twitter subnetwork. In contrast, our algorithm can provide  (pseudo) \emph{ground truth} with  guaranteed small (relative) error (e.g. 0.5\%). Thus it is a more concrete benchmark tool for research on network cascades.
\end{itemize}

\textbf{Organization.} The rest of the paper is organized as follows: In Section~\ref{sec:model}, we introduce the diffusion model and the definition of outward influence with its properties. We propose an \FPRAS{} for outward influence estimation in Section~\ref{sec:outest}. Applications in influence estimation are presented in Section~\ref{sec:inf} which is followed by the experimental results in Section~\ref{sec:exp} and conclusion in Section~\ref{sec:con}. We cover the most recent related work in Section~\ref{sec:work}.

%% file: body/model.tex
\vspace{-0.05in}
\section{Definitions and Properties}
\label{sec:model}

In this section, we will introduce stochastic diffusion models, the new measure of \emph{Outward Influence}, and showcase its properties under the popular Independent Cascade (IC) model \cite{Kempe03}. 

\textbf{Diffusion model.} Consider a network abstracted as a graph $\G=(V,E)$, where $V$ and $E$ are the sets of nodes and edges, respectively.  For example, in a social network, $V$ and $E$ correspond to the set of users and their social relationships, respectively. Assume that there is a cascade starting from a subset of nodes $S \subseteq V$, called \emph{seed set}. How the cascade progress is described by a diffusion model (aka cascade model) $\mathcal{M}$ that dictates how nodes gets activated/influenced. In a stochastic diffusion model, the cascade is dictated by a random vector $\theta$ in a sample space $\Omega_\theta$. Describing the diffusion model is then equivalent to specifying the distribution $P$ of $\theta$.

 Let $r_S(\theta)$ be the size of the cascade, the number of activated nodes in the end. The influence spread of $S$, denoted by $\I(S)$,  under diffusion model $\mathcal{M}$ is the expected size of the cascade, i.e.,
\vspace{-0.05in}
\begin{align}
	\label{eq:inf_gen}
	\I(S) =
	\begin{cases}
	\sum_{\theta \in \Omega_\theta} r_\theta(S) \Pr[\theta] \text{ \ \ \ for discrete } \Omega_\theta,\\
	\int_{\theta \in \Omega_\theta} r_\theta(S) d P(\theta)  \text{ \ \ \ \ for continuous } \Omega_\theta
	\end{cases}
\end{align}
\vspace{-0.1in}

\noindent For example, we describe below  the unknown vector $\theta$ and their distribution for the most popular diffusion models.

\begin{itemize}
	\item Information diffusion models, e.g. Independent Cascade (IC), Linear Threshold (LT), the general triggering model \cite{Kempe03}:  $\theta \in \{0,1\}^{|E|}$, and $\forall (u,v) \in E, \theta_{(u,v)}$ is a Bernouli random variable that indicates whether $u$ activates/influences $v$. That is for given $w(u,v) \in (0, 1)$, $\theta(u, v)=1$ if $u$ activates $v$ with a probability $w(u, v)$ and 0, otherwise. 
	\item Epidemic cascading models, e.g., Susceptible-Infected (SI) \cite{Daley01, Nguyen165} and its variations:  $\theta \in \mathbb{N}^{|E|}$, and $\forall (u,v) \in E, \theta_{(u,v)}$ is a random variable following a geometric distribution. $\theta_{(u, v)}$ indicates how long it takes $u$ to activates $v$ after $u$ is activated.
	\item Continuous-time models \cite{Du13}:  $\theta \in \mathbb{R}^{|E|}$, and $\theta_{(u, v)}$ is a continuous random variable with density function  $\pi_{u, v}(t)$. $\theta_{(u,v)}$ also indicates the transmission times (time until $u$ activates $v$) like that in the SI model, however, the transmissions time on different edges follow different distributions.
\end{itemize}

\textbf{Outward Influence.} We introduce the notion of \textsf{Outward Influence} which captures the influence of a subset of nodes towards the rest of the network. Outward influence excludes the self-influence of the seed nodes from the measure. 

\begin{Definition}[Outward Influence]
	Given a graph $\G = (V,E)$, a set $S \subseteq V$ and a diffusion model $\mathcal{M}$, the Outward Influence of $S$, denoted by $\Ext(S)$, is
	\vspace{-0.05in}
	\begin{align}
		\label{eq:def}
		\Ext(S) = \I(S) - |S|
	\end{align}
	%
\end{Definition}
Thus, influence and outward influence of a seed set $S$  differ exactly by the number of nodes in $S$.

\textbf{Influence Spread/Outward Influence Estimations.} A fundemental task in network science is to estimate the influence of a given seed set $S$. Since the exact computation is \#P-hard (Subsection 2.2), we aim for estimation with bounded error.

\begin{Definition}[Influence Spread Estimation] \text{ \ \ \ \ \ \ \ \ \  } Given a graph $\G$ and a set $S \subseteq V$, the problem asks for an $(\epsilon,\delta)$-estimate $\hat \I(S)$ of influence spread $\I(S)$, i.e.,
	\vspace{-0.05in}
	\begin{equation}
	\Pr[(1-\epsilon)\I(S) \leq \hat \I(S) \leq (1+\epsilon)\I(S)] \geq 1-\delta.
	\end{equation}
\end{Definition}
The outward influence estimation problem is stated similarly:
\begin{Definition}[Outward Influence  Estimation] \text{ \ \ \ \ \ \ \ \ \ \ } Given a graph $\G$ and a set $S \subseteq V$, the problem asks for an $(\epsilon,\delta)$-estimate $\hExt(S)$ of influence spread $\Ext(S)$, i.e.,
	\vspace{-0.05in}
	\begin{equation}
	\Pr[(1-\epsilon)\Ext(S) \leq \hExt(S) \leq (1+\epsilon)\Ext(S)] \geq 1-\delta.
	\end{equation}
\end{Definition}
A common approach for estimation is through generating independent Monte-Carlo samples and taking the average. However, one faces two major challenges:
\begin{itemize}
	\item How to achieve a minimum number samples to get an $(\epsilon, \delta)$-approximate?  
	\item How to effectively generate samples with small variance, and, thus, reduce the number of samples?
\end{itemize}


For simplicity, we focus on the well-known  \emph{Independent Cascade} (IC) model and provide the extension of our approaches to other cascade models in Subsection~\ref{subsec:extension}.


%

\vspace{-0.1in}
\subsection{Independent Cascade (IC) Model}
Given a probabilistic graph $\G = (V, E)$ in which each edge $(u,v) \in E$ is associated with a number $w(u,v) \in (0, 1)$. $w(u,v)$ indicates the probability that node $u$ will successfully activate $v$ once $u$ is activated. In practice, the probability $w(u,v)$ can be mined from interaction frequency \cite{Kempe03,Tang14} or learned from action logs \cite{Goyal10}. 

\textbf{Cascading Process.} The cascade starts from a subset of nodes $S \subseteq V$, called seed set. The cascade happens in discrete rounds $t=0, 1,...|V|$. At round $0$, only nodes in $S$ are active and the others are inactive. When a node $u$ becomes active, it has a single chance to activate (aka influence) each neighbor $v$ of $u$ with probability $w(u,v)$. An active node remains active till the end of the cascade process. It stops when no more nodes get activated.

\textbf{Sample Graph.} Associate with each edge $(u,v) \in E$ a biased coin that lands heads with probability $w(u,v)$ and tails with probability $1-w(u,v)$. Deciding the outcome when $u$ attempts to activate $v$ is then equivalent to the outcome of flipping the coin.  If the coin landed heads, the activation attemp succeeds and we call $(u,v)$ a \textit{live-edge}. Since all the activation on the edges are independent in the IC model, it does not matter when we flip the coin. That is we can flip all the coins associated with the edges $(u, v)$  at the same time instead of waiting until node $u$ becomes active. We call the graph $g$ that contains the nodes $V$ and all the live-edges a \textit{sample graph} of $\mathcal G$.

Note that the model parameter $\theta$ for the IC is a random vector indicating the states of the edges, i.e. \textit{live-edge} or not. In other words, $\Omega_\theta$ corresponds to the space of all possible sample graphs of $\G$, denoted by $\Omega_\G$.

\textbf{Probabilistic Space.} The graph $\mathcal G$ can be seen as a generative model. The set of all sample graphs generated from $\mathcal G$ together with their probabilities define a probabilistic space $\Omega_\G$. Recall that each sample graph $g \in \Omega_\G$ can be generated by flipping coins on all the edges to determine whether or not the edge is live or appears in $g$. Each edge $(u, v)$ will be present in the a sample graph with probability $w(u, v)$. Thus, the probability that a sample graph $g=(V, E' \subseteq E)$ is generated from $\G$ is
\vspace{-0.05in}
\begin{align}
	\Pr[g \sim \G] = \prod_{(u,v) \in E'} w(u,v) \prod_{(u,v) \in E \backslash  E'} (1-w(u,v)).
\end{align}
\vspace{-0.1in}

{\bf Influence Spread and Outward Influence.} In a sample graph $g \in \Omega_\G$, let $r_g(S)$ be the set of nodes reachable from $S$. The \emph{influence spread} in Eq.~\ref{eq:inf_gen} is rewritten,
\vspace{-0.05in}
\begin{align}
	\label{eq:inf}
	\I(S) = \sum_{g \in \Omega_\G} |r_g(S)| \Pr[g \sim \G],
\end{align}
\vspace{-0.1in}

\noindent and the outward influence is defined accordingly to Eq.~\ref{eq:def},
\vspace{-0.05in}
\begin{align}
	\Ext(S) = \I(S) - |S|
\end{align}
\vspace{-0.25in}

%% file: body/hardness.tex
\subsection{Outward Influence under IC}
We show the properties of outward influence under the IC model.

\textbf{Better Influence Discrepancy.} As illustrated through Fig. \ref{fig:example}, the elimination of the nominal constant $|S|$ helps to differentiate the ``actual influence'' of the seed nodes to the other nodes in the network. In the extreme case when $p = o(1)$, the ratio between the influence spread of $u$ and $v$ is $\frac{1+p+2p^2}{1+p+2p} \approx 1$, suggesting $u$ and $v$ have the same influence. However, outward influence can capture the fact that $v$ can influence roughly twice the number of nodes than $u$, since s $\frac{\Ext(u)}{\Ext(v)} = \frac{p+2p^2}{2p} \approx 1/2$.

\textbf{Non-monotonicity.} Outward influence as a function of seed set $S$ is non-monotone. This is different from the influence spread.
In Figure~\ref{fig:example}, $\Ext(\{u\}) = 0.12 < \Ext(\{u,v\}) = 0.2$, however,  $\Ext(\{u\}) = 0.12 > \Ext(\{u,w\}) = 0.11$. That is adding nodes to the seed set may increase or decrease the outward influence.

\textbf{Submodularity.} A submodular function expresses the diminishing returns behavior of set functions and are suitable for many applications, including approximation algorithms and machine learning.
If $\Omega$ is a finite set, a submodular function is a set function $f : 2^\Omega \leftarrow \mathbb{R}$, where $2^\Omega$ denotes the power set of $\Omega$, which satisfies that for every $X,Y\subseteq \Omega$ with $X \subseteq Y$ and every $x\in \Omega \setminus Y$, we have,
\vspace{-0.05in}
\begin{align}
f(X\cup \{x\})-f(X)\geq f(Y\cup \{x\})-f(Y).
\end{align}
\vspace{-0.2in}

Similar to influence spread, outward influence, as a function of the seed set $S$, is also submodular.
\begin{Lemma}
	\label{lem:submod}
	Given a network $G = (V,E,w)$, the outward influence function $\Ext(S)$ for $S \in 2^{|V|},$ is a submodular function 
\end{Lemma}

\vspace{-0.1in}
\subsection{Hardness of Computation} 
\label{subsec:hardness}
If we can compute outward influence of $S$, the influence spread of $S$ can be obtained by adding $|S|$ to it. Since computing influence spread is \#P-hard \cite{Chen10}, it is no surprise that computing outward influence is also \#P-hard. 
\begin{Lemma}
	Given a probabilistic graph $G = (V,E,w)$ and a seed set $S \subseteq V$, it is \#P-hard to compute $\Ext(S)$.
\end{Lemma}


However, while influence spread is lower-bounded by one, the outward influence of any set $S$ can be arbitrarily small (or even zero). Take an example in Figure~\ref{fig:example}, node $u$ has influence of $\I(\{u\}) = 1+p +2p^2\geq 1$ for any value of $p$. However, $u$'s outward influence $\Ext(\{u\}) = p+2p^2$ can be exponentially small if $p= \frac{1}{2^n}$. This makes estimating outward influence challenging, as the number of samples needed to estimate the mean of random variables is inversely proportional to the mean.

%

\textbf{Monte-Carlo estimation}. A typical approach to obtain an $(\epsilon,\delta)$-approximaion of a random variable is through Monte-Carlo estimation: taking the average over many samples of that random variable. Through the Bernstein's inequality \cite{Dagum00}, we have the lemma:
\begin{Lemma}
	\label{lem:mc}
	Given a set $X_1, X_2, \dots$ of i.i.d. random variables having a common mean $\mu_X$, there exists a Monte-Carlo estimation which gives an $(\epsilon,\delta)$-approximate of the mean $\mu_X$ and uses $T = O(\frac{1}{\epsilon^2} \ln (\frac{2}{\delta}) \frac{b}{\mu_X})$ random variables where $b$ is an upper-bound of $X_i$, i.e. $X_i \leq b$.
\end{Lemma}

To estimate the influence spread $\I(S)$, existing work often simulates the cascade process using a BFS-like procedure and takes the average of the cascades' sizes as the influence spread. The number of samples needed to obtain an $(\epsilon, \delta)$-approximation is $O(\frac{1}{\epsilon^{2}}\log{\left(\frac{1}{\delta}\right)}\frac{n}{\I(S)})$ samples. Since $\I(S) \geq 1$, in the worst-case, we need only a polynomial number of samples, $O(\frac{1}{\epsilon^{2}}\log{\left(\frac{1}{\delta}\right)}n)$. 

Unfortunately, the same argument does not apply for the case of $\Ext(S)$, since $\Ext(S)$ can be arbitrarily close to zero. For the same reason, the recent advances in influence estimation in \cite{Borgs14, Lucier15} cannot be adapted to obtain a polynomial-time algorithm to compute an $(\epsilon, \delta)$-approximation (aka \FPRAS) for outward influence. We shall address this challenging task  in the next section.

We summarize the frequently used notations in Table~\ref{tab:syms}.
\renewcommand{\arraystretch}{1.1}

\setlength\tabcolsep{3pt}
\begin{table}[hbt]\small
	\centering
	\caption{Table of notations}
	\vspace{-0.15in}
	\begin{tabular}{m{1.1cm}|m{6.5cm}}
		\addlinespace
		\toprule
		\bf Notation  &  \quad \quad \quad \bf Description \\
		\midrule 
		$n, m$ & \#nodes, \#edges of graph $\G=(V, E, w)$.\\
		\hline
		$\I(S)$ & Influence Spread of seed set $S\subseteq V$.\\
		\hline
		$\Ext(S)$ & Outward Influence of seed set $S\subseteq V$.\\
		\hline
		$N^{out}(u)$ & The set of out-neighbors of $u$: $N^{out}(u)=\{ v \in V | (u, v) \in E \}$\\
		\hline
		$N^{out}_S$ & $N^{out}_S = \bigcup_{u \in S} N^{out}(u)\backslash S$.\\
		\hline
		$A_i$ & The event that $v_i$ is active and $v_1, \dots, v_{i-1}$ are not active after round 1. \\
		\hline
		$\beta_0$ & $\beta_0 = \sum_{i = 1}^{l} \Pr[A_i] = 1-\Pr[A_{l+1}]$.\\
		\hline
		$c(\epsilon,\delta)$ & $c(\epsilon,\delta) = (2+\frac{2}{3}\epsilon)\ln(\frac{2}{\delta})\frac{1}{\epsilon^2}$ \\
		\hline
		$\epsilon'$ & $\epsilon' = \epsilon \Big(1-\frac{\epsilon b}{(2+\frac{2}{3}\epsilon)\ln(\frac{2}{\delta})(b-a)}\Big) \approx \epsilon (1- O(\frac{1}{\ln n}))$ for $\delta=\frac{1}{n}$ \\
		\hline
		$\Upsilon$ & $\Upsilon = (1+\epsilon)c(\epsilon',\delta) (b-a)$\\
		\bottomrule
	\end{tabular}%
	\label{tab:syms}%
	\vspace{-0.05in}
\end{table}%

%
%
%
%
%
%
%
%

%% file: body/fpras.tex
\section{Outward Influence Estimation via Importance Sampling}
\label{sec:outest}


We propose a Fully Polynomial Randomized Approximation Scheme (\FPRAS{}) to estimate the outward influence of a given set $S$. Given two precision parameters $\epsilon,\delta \in (0, 1)$, our \FPRAS{} algorithm guarantees to return an $(\epsilon,\delta)$-approximate $\hExt(S)$ of the outward influence $\Ext(S)$,
\vspace{-0.05in}
\begin{align}
	\Pr[(1-\epsilon)\Ext(S) \leq \hExt(S) \leq (1+\epsilon)\Ext(S)] \geq 1-\delta.
\end{align}
\textbf{General idea.} 
Our starting point is an observation that the cascade triggered by the seed set with small influence spread often stops right at round $0$. The probability of such cascades, termed \emph{trivial cascades}, can be computed exactly. Thus if we can sample only the \emph{non-trivial cascades}, we will obtain a better sampling method to estimate the outward influence. The reason is that the ``outward influence'' associated with non-trivial cascade is also lower-bounded by one. Thus, we again can apply the argument in the previous section on the polynomial number of samples.

Given a graph $\G$ and a seed set $S$, we introduce our \emph{importance sampling} strategy to generate such non-trivial cascades. It consists of two stages: 
\begin{enumerate} 
\item Guarantee that at least one neighbor of $S$ will be activated through a biased selection towards the cascades with at least one node outside of $S$ and,
\item Continue to simulate the cascade using the standard procedure following the diffusion model.
\end{enumerate}
This importance sampling strategy is general for different diffusion models. In the following, we illustrate our importance sampling under the focused IC model.

\subsection{Importance IC Polling}
We propose \textit{Importance IC Polling} (\IICP{}) to sample non-trivial cascades in Algorithm~\ref{alg:simulate}. 
\begin{figure}[h!] \centering
	\vspace{-0.02in}
	\includegraphics[width=0.4\linewidth]{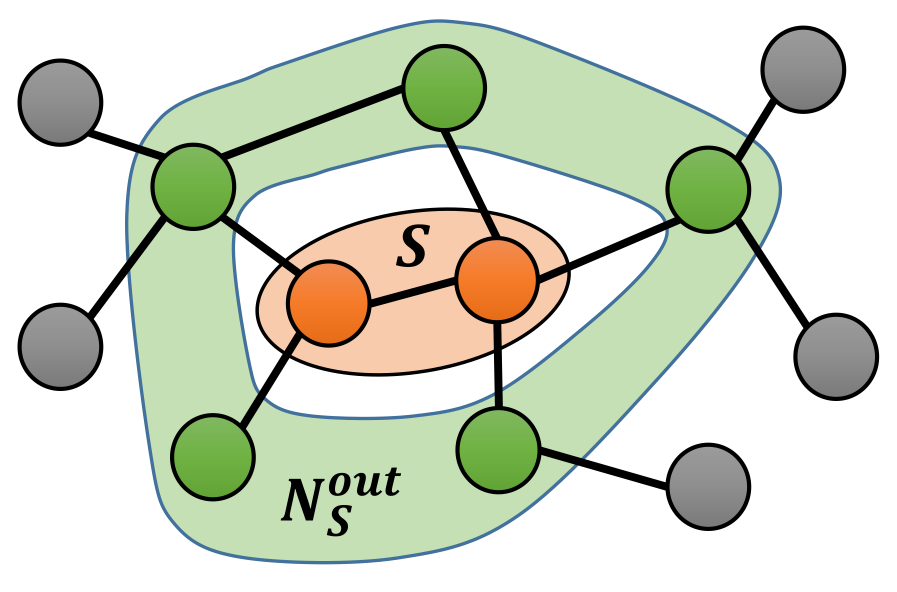}
	\vspace{-0.1in}
	\caption{Neighbors of nodes in $S$}
	\label{fig:rr}
	\vspace{-0.05in}
\end{figure}

First, we ``merge'' all the nodes in $S$ and define a ``unified neighborhood'' of $S$. Specifically, let  $N^{out}(u) = \{ v | (u, v) \in E \}$ the set of out-neighbors of $u$ and 
 $N^{out}_S=\bigcup\limits_{u \in S}N^{out}_u \backslash S$ the set of out-neighbors of $S$ \emph{excluding $S$}. For each $v \in N^{out}_S$,
 \vspace{-0.05in}
\begin{equation}
	\label{eq:psv}
	P_{S,v} = 1 - \prod_{u \in S}(1-w(u,v)),
\end{equation} 
\vspace{-0.1in}

\noindent the probability that $v$ is activated directly by one (or more) node(s) in $S$. Without loss of generality, assume that $P_{S,v} \le 1$ (otherwise, we simply add $v$ into $S$).
 
Assume an order on the neighborhood of $S$, that is
\vspace{-0.05in}
 \[
	N^{out}_S = \{v_1, v_2, \dots, v_l \},
 \] where $l=|N^{out}_S|$.  
For each $i=1..l$, let $A_i$ be the event that $v_i$ be the ``first'' node that gets activated directly by $S$:
\[
A_i=\{v_1,\ldots, v_{i-1}\ \text{are not active and } v_i  \text{ is active after round }1\}.
\]
The probability of  $A_i$ is
\vspace{-0.1in}
\begin{equation}
	\label{eq:pai}
	\Pr[A_i] = P_{S,v_i}\prod_{j=1}^{i-1}(1-P_{S,v_j}).
\end{equation}
\vspace{-0.1in}

\noindent For consistency, we also denote $A_{l+1}$ the event that none of the neighbors are activated, i.e.,
\vspace{-0.1in}
\begin{align}
	\label{eq:pal}
	\Pr[A_{l+1}] = 1 - \sum_{i = 1}^{l} \Pr[A_i].
\end{align}
\vspace{-0.1in}

\noindent Note that $A_{l+1}$ is also the event that the cascade stops right at round $0$. Such a cascade is termed a \emph{trivial cascade}. As we can compute exactly the probability of trivial cascades, we do not need to sample those cascades but focus only on the non-trivial ones.

Denote by $\beta_0$  the probability of having at least one nodes among $v_1, \ldots,v_l$ activated by $S$, i.e., 
\vspace{-0.1in}
\begin{align}
\label{eq:zerop}
\beta_0 = \sum_{i = 1}^{l} \Pr[A_i] = 1-\Pr[A_{l+1}].
\end{align}
\vspace{-0.15in}

\let\oldnl\nl
\newcommand{\nonl}{\renewcommand{\nl}{\let\nl\oldnl}}

\begin{algorithm}[!h] \small
	\caption{\IICP{} - Importance IC Polling}
	\label{alg:simulate}
	\KwIn{A graph $\G=(V,E,w)$ and a seed set $S$}
	\KwOut{$Y^{(S)}$ - size of a random outward cascade from $S$}
	\vspace{0.05in}
	\nonl \textbf{Stage 1 } \tcp{Sample non-trivial neighbors of set S}
	Precompute $\Pr[A_i], i = 1, \dots, l+1$ using Eq.~\ref{eq:pai} and Eq.~\ref{eq:pal}\\
	Select one neighbor $v_i$ among $v_1, \dots, v_l$ with probability of selecting $v_i$ being $\frac{\Pr[A_i]}{\beta_0}$\\
	Queue $R \leftarrow \{v_i\}; Y^{(S)} = 1$; Mark $v_i$ and all nodes in $S$ visited\\
	\For{$j = i+1 : l$}{
		With a probability $P_{S,v_j}$ do\\
		\hspace{0.2in}	Add $v_j$ into $R$; $Y^{(S)} \leftarrow Y^{(S)} + 1$; Mark $v_j$ visited. 
	}
	\vspace{0.05in}
	\nonl \textbf{Stage 2 } \tcp{Sample from newly influenced nodes}
	\While{$R$ is non-empty}{
		$u \leftarrow R$\textsf{.pop()}\\
		\ForEach{unvisited neighbor $v$ of $u$}{
			With a probability $w(u, v)$\\				
			\hspace{0.2in}Add $v$ to $R$; $Y^{(S)} \leftarrow Y^{(S)}+1$; Mark $v$ visited.
		}
	}
	\textbf{return} $Y^{(S)}$;\\
\end{algorithm}
We now explain the details in the Importance IC Polling Algorithm (\IICP{}), summarized in Alg.~\ref{alg:simulate}. The algorithm outputs the size of the cascade minus the seed set size. We term the output of \IICP{}  the \emph{outer size} of the cascade. The algorithm consists of two stages. 

\textbf{Stage 1}.
By definition, the events $A_i, A_2, ..., A_l, A_{l+1}$ are disjoint and form a partition of the sample space. To generate a non-trivial cascade, we first select in the first round $v_i, i = 1, \dots, l$ with a probability $\frac{\Pr[A_i]}{\beta_0}, i = 1,\dots,l$ (excluding $A_{l+1}$). This will guarantee that at least one of the neighbors of $S$ will be activated. Let $v_i$ be the selected node,  after the first round $v_i$ becomes active and $v_1,\ldots, v_{i-1}$ remains inactive. The nodes $v_j$ among $v_{i+1}, \dots, v_l$ are then activated independently with probability $P_{S,v_j}$ (Eq.~\ref{eq:psv}). 

\textbf{Stage 2.}
After the first stage of sampling neighbors of $S$, we get a non-trivial set of nodes directly influenced from $S$. For each of those nodes and later influenced nodes, we will sample a set of its neighbors by the naive BFS-like IC polling scheme \cite{Kempe03}. Assume sampling neighbors of a newly influenced node $u$, each neighbor $v_j \in N^{out}(u)$ is influenced by $u$ with probability $w(u,v_j)$. The neighbors of those influenced nodes are next to be sampled in the same fashion.

In addition, we keep track of the newly influenced nodes using a queue $R$ and the number of active nodes outside $S$ using $\Ys{}$.

The following lemma shows how to estimate the (expected) cascade size through the (expected) outer size of non-trivial cascades.
\begin{Lemma}
	\label{lem:sim_prob}
	Given a seed set $S \subseteq V$, let $Y^{(S)}$ be the random variable associated with the output of the \IICP{} algorithm. The following properties hold,
	\begin{itemize}
		\item $1 \leq Y^{(S)} \leq n - |S|,$
		\item $\Ext(S) = \E[Y^{(S)}] \cdot \beta_0.$
	\end{itemize}
\end{Lemma}

Further, let $\Omega_W$ be the probability space of non-trivial cascades and $\Omega_Y$ the probability space for the outer size of non-trivial cascades, i.e, $\Ys{}$. The probability of $\Ys{} \in [1, n-|S|]$ is given by,
\vspace{-0.1in}
\begin{align}
	\Pr[\Ys{} \in \Omega_Y] = \sum_{\Ws{} \in \Omega_W, |\Ws{}| = \Ys{}} \Pr[\Ws{} \in \Omega_W]. \nonumber
\end{align}

\subsection{FPRAS for Outward Influence Estimation}

From Lemma~\ref{lem:sim_prob}, we can obtain an estimate $\hExt(S)$ of $\Ext(S)$ through getting an estimate  $\hat \E[\Ys{}]$ of $\E[\Ys{}]$ by,
\vspace{-0.05in}
\begin{align}
	& \Pr\Big [(1-\epsilon)\E[\Ys{}] \leq \hat \E[\Ys{}] \leq (1+\epsilon)\E[\Ys{}] \Big ] \nonumber \\
	& = \Pr\Big [(1-\epsilon)\E[\Ys{}]\beta_0 \leq \hat \E[\Ys{}]\beta_0 \leq (1+\epsilon)\E[\Ys{}]\beta_0 \Big ] \nonumber \\
	& = \Pr \Big [(1-\epsilon)\Ext(S) \leq \hExt(S) \leq (1+\epsilon)\Ext(S) \Big ],
\end{align}
\vspace{-0.15in}

\noindent where the estimate $\hExt(S) = \hat \E[\Ys{}]\cdot \beta_0$. Thus, finding an $(\epsilon,\delta)$-approximation of $\Ext(S)$ is then equivalent to finding an $(\epsilon,\delta)$-approximate $\hat \E[\Ys{}]$ of $\E[\Ys{}]$.

The advantage of this approach is that estimating $\E[\Ys{}]$, in which the random variable $\Ys{}$ has value of at least $1$, requires only a polynomial number of samples. Here the same argument on the number of samples to estimate influence spread in subsection~\ref{subsec:hardness} can be applied. Let $\Ys_1, \Ys_2, \dots$ be the random variables denoting the output of \IICP.  We can apply Lemma~\ref{lem:mc} on the set of random variables $\Ys_1, \Ys_2, \dots$ satisfying $1 \leq \Ys_i \leq |V| - |S|$. Since each random variable $\Ys_i$ is at least 1 and hence, $\mu_Y = \E[\Ys{}] \geq 1$, we need at most a polynomial $T = O(\ln(\frac{2}{\delta}) \frac{1}{\epsilon^2} (n-|S|))$ random variables for the Monte-Carlo estimation. Since, \IICP{} has a worst-case time complexity $O(m+n)$, the Monte-Carlo using \IICP{} is an \FPRAS{} for estimating outward influence.
\begin{theorem}
 Given arbitrary $0 \leq \epsilon,\delta \leq 1$ and a set $S$, the Monte-Carlo estimation using \IICP{} returns an $(\epsilon,\delta)$-approximation of $\Ext(S)$ using $O(\ln(\frac{2}{\delta}) \frac{1}{\epsilon^2} (n-|S|))$ samples.
\end{theorem}

In Section \ref{sec:inf}, we will show that both outward influence and influence spread can be estimated by a powerful algorithm saving a factor of more than $\frac{1}{\epsilon}$ random variables compared to this \FPRAS{} estimation. The algorithm is built upon our mean estimation algorithms for bounded random variables proposed in the following.

%% file: body/mean_est.tex
\section{Efficient Mean Estimation for\\ Bounded Random Variables}
\label{sec:mean}
In this section, we propose an efficient mean estimation algorithm for bounded random variables. This is the core of our algorithms for accurately and efficiently estimating the outward influence and influence spread in Section~\ref{sec:inf}.

We first propose an `intermediate' algorithm: \textit{Generalized Stopping Rule Estimation} (\GSRA) which relies on a simple stopping rule and returns an $(\epsilon,\delta)$-approximate of the mean of lower-bounded random variables. The \GSRA{} simultaneously generalizes and fixes the error of the Stopping Rule Algorithm \cite{Dagum00} which only aims to estimate the mean of $[0,1]$ random variables and has a technical error in its proof.

The main mean estimation algorithm, namely Robust Sampling Algorithm (\SiEIE{}) presented in Alg.~\ref{alg:extinf}, effectively takes into account both mean and variance of the random variables. It uses \GSRA{} as a subroutine to estimate the mean value and variance at different granularity levels.

\subsection{Generalized Stopping Rule Algorithm}	
We aim at obtaining an $(\epsilon, \delta)$-approximate of the mean of random variables $X_1, X_2,\dots$. Specifically, the random variables are required to satisfy the following conditions:
\begin{itemize}
	\item $a \leq X_i \leq b$, $\forall i=1, 2, \dots$
	\item $\E[ X_{i+1} | X_1, X_2,..., X_i ] = \mu_X,$ $\forall i=1, 2,\dots$
\end{itemize}
where $0\leq a < b$ are fixed constants and (unknown) $\mu_X$. 

Our algorithm generalizes the stopping rule estimation in \cite{Dagum00} that provides $(\epsilon, \delta)$ estimation of the mean of i.i.d. random variables $X_1, X_2,...  \in [0, 1]$. The notable differences are the following:
\begin{itemize}
	\item We discover and amend an error in the stopping algorithm in \cite{Dagum00}: the number of samples drawn by that algorithm may not be sufficient to guarantee the $(\epsilon, \delta)$-approximation. 
	\item We allow estimating the mean of random variables that are \emph{possibly dependent} and/or with \emph{different distributions}. Our algorithm works as long as the random variables have the same means. In contrast, the algorithm in \cite{Dagum00} can only be applied for i.i.d random variables.
	\item Our proposed algorithm obtains an unbiased estimator of the mean, i.e. $\E[\hat \mu_X] = \mu_X$ while \cite{Dagum00} returns a biased one.
	\item Our algorithm is faster than the one in \cite{Dagum00} whenever the lower-bound for random variables $a >0$. 	
\end{itemize}


\begin{algorithm} \small
	\caption{Generalized Stopping Rule Alg. (\GSRA{})}
	\label{alg:stop}
	\KwIn{Random variables $X_1, X_2, \dots$ and $0 < \epsilon,\delta < 1$}
	\KwOut{An $(\epsilon,\delta)$-approximate of $\mu_X = E[X_i] $}
	If $b-a<\epsilon b$, \textbf{return} $\mu_X = a$.\\
	Compute: $\epsilon' = \epsilon \Big(1-\frac{\epsilon b}{(2+\frac{2}{3}\epsilon)\ln(\frac{2}{\delta})(b-a)}\Big);\Upsilon = (1+\epsilon)c(\epsilon',\delta) (b-a)$;\\
	Initialize $h = 0, T = 0$;\\
	\While{$h < \Upsilon$}{
		$ h \leftarrow h + X_T, T \leftarrow T+1$;\\
	}
	\textbf{return} $\hat \mu_X = h/T$;
\end{algorithm}

Our Generalized Stopping Rule Algorithm (\GSRA{}) is described in details in Alg.~\ref{alg:stop}. Denote $c(\epsilon,\delta) = (2+\frac{2}{3}\epsilon)\ln(\frac{2}{\delta})\frac{1}{\epsilon^2}$.

The algorithm contains two main steps: 1) Compute the stopping threshold $\Upsilon$ (Line~2) which relies on the value of $\epsilon'$ computed from the given precision parameters $\epsilon,\delta$ and the range $[a,b]$ of the random variables; 2) Consecutively acquire the random variables until the sum of their outcomes exceeds $\Upsilon$ (Line~4-5). Finally, it returns the average of the outcomes, $\hat \mu_X = h/T$ (Line~6), as an estimate for the mean, $\mu_X$. Notice that $\Upsilon$ in \GSRA{} depends on $(b-a)$ and thus, getting tighter bounds on the range of random variables holds a key for the efficiency of \GSRA{} in application perspectives.

The approximation guarantee and number of necessary samples are stated in the following theorem.
\begin{theorem}
	\label{theo:stop}
	The Generalized Stopping Rule Algorithm (\GSRA{}) returns an $(\epsilon,\delta)$-approximate $\hat \mu_X$ of $\mu_X$, i.e., 
	\begin{align}
		\Pr[(1-\epsilon)\mu_X \leq \hat \mu_X \leq (1+\epsilon)\mu_X] > 1- \delta,
	\end{align}
	and, the  number of samples $T$ satisfies,
	\begin{align}
		\Pr[T \leq (1+\epsilon)\Upsilon/\mu_X] > 1 - \delta/2.
	\end{align}
\end{theorem}

\textbf{The hole in the Stopping Rule Algorithm in \cite{Dagum00}.} The estimation algorithm in \cite{Dagum00} for estimating the mean of random variables in range $[0,1]$ also bases on a main stopping rule condition as our \GSRA{}. It computes a threshold
\vspace{-0.15in}
\begin{align}
	\Upsilon_1 = 1+ (1+\epsilon) 4 (e-2) \ln (\frac{2}{\delta}) \frac{1}{\epsilon^2},
\end{align}
\vspace{-0.15in}

\noindent where $e$ is the base of natural logarithm, and generates samples $X_j$ until $\sum_{j = 1}^{T}X_j \geq \Upsilon_1$. The algorithm returns $\hat \mu_X = \frac{\Upsilon_1}{T}$ as a biased estimate of $\mu_X$.

Unfortunately, the threshold $\Upsilon_1$ to determine the stopping time does not completely account for the fact that the necessary number of samples should go over the expected one in order to provide high solution guarantees. This actually causes a flaw in their later proof of the correctness. 

To amend the algorithm, we slightly strengthen the stopping condition by replacing the $\epsilon$ in the formula of $\Upsilon$ with an $\epsilon' = \epsilon \Big(1-\frac{\epsilon b}{(2+\frac{2}{3}\epsilon)\ln(\frac{2}{\delta})(b-a)}\Big)$ (Line 2, Alg.~\ref{alg:stop}). Since $\epsilon b < b-a$ (else the algorithm returns $\mu_X = a$) and assume w.l.o.g. that $\delta < 1/2$, it follows that $\epsilon' \geq 0.729 \epsilon$. Thus the number of samples, in comparison to those in the stopping rule algorithm in \cite{Dagum00} increases by at most a constant factor.

\textbf{Benefit of considering the lower-bound $a$.} By dividing the random variables by $b$, one can apply the stopping rule algorithm in \cite{Dagum00} on the normalized random variables. The corresponding value of $\Upsilon$ is then 
\vspace{-0.05in}
\begin{align}
	\Upsilon = 1+ (1+\epsilon) (2+\frac{2}{3}\epsilon) \ln (\frac{2}{\delta}) \frac{1}{\epsilon'^2} b
\end{align}
$\Upsilon$ in our proposed algorithm is however smaller by a multiplicative factor of $\frac{b-a}{b}$. Thus it is faster than the algorithm in \cite{Dagum00} by a factor of $\frac{b-a}{b}$ on average. Note that in case of estimating the influence, we have $a = 1, b = n - |S|$. Compared to algorithm applied \cite{Dagum00} directly, our \GSRA{} algorithm saves the generated samples by a factor of $\frac{b-a}{b} = \frac{n-|S|-1}{n} = 1 - \frac{|S|+1}{n}< 1$.

\textbf{Martingale theory to cope with weakly-dependent random variables.} To prove Theorem~\ref{theo:stop}, we need a stronger Chernoff-like bound to deal with the general random variables $X_1, X_2, \dots$ in range $[a,b]$ presented in the following.

Let define random variables $Y_i = \sum_{j = 1}^{i} (X_j - \mu_X), \forall i \geq 1$. Hence, the random variables $Y_1, Y_2, \dots$ form a Martingale \cite{Mitzenmacher05} due to the following,
\begin{align}
	\E[Y_i | Y_1, \dots, Y_{i-1}] = \E[Y_{i-1}] + \E[X_i - \mu_X] = \E[Y_{i-1}]. \nonumber
\end{align}

Then, we can apply the following lemma from \cite{Chung06} stating,
\begin{Lemma}
	\label{lem:martingale}
	Let $Y_1, \dots, Y_i,...$ be a martingale, such that $|Y_1| \leq \alpha$, $|Y_j - Y_{j-1}| \leq \alpha$ for all $j = [2,i]$, and
	\vspace{-0.1in}
	\begin{align}
		\Var[Y_1] + \sum_{j = 2}^i \Var[Y_j | Y_1, \dots, Y_{j - 1}] \leq \beta.
	\end{align}
	\vspace{-0.15in}
	
	\noindent Then, for any $\lambda \geq 0$,
	\vspace{-0.1in}
	\begin{align}
		\Pr[Y_i - \E[Y_i] \geq \lambda] \leq \exp(-\frac{\lambda^2}{2/3\cdot \alpha \cdot \lambda + 2 \cdot \beta})
	\end{align}
	\vspace{-0.15in}
\end{Lemma}
In our case, we have $|Y_1| = |X_1 - \mu_X| \leq b - a$, $|Y_j - Y_{j - 1}| = |X_i - \mu_X| \leq b - a$, $\Var[Y_1] = \Var[X_1 - \mu_X] = \Var[X]$ and $\Var[Y_j | Y_1, \dots, Y_j] = \Var[X_j - \mu_X] = \Var[X]$. Apply Lemma~2 with $i = T$ and $\lambda = \epsilon T \mu_X$, we have,
\vspace{-0.05in}
\begin{align}
	&\Pr\Big[\sum_{j = 1}^{T} X_j \geq (1+\epsilon)\mu_X T\Big] \leq \exp \big(\frac{- \epsilon^2 T^2 \mu^2_X}{\frac{2}{3}(b-a)\epsilon\mu_X T + 2 \Var[X] T} \big)
	\label{eq:eq33}
\end{align}
\vspace{-0.1in}

Then, since $\Var[X] \leq \mu_X (b - \mu_X) \leq \mu_X(b-a)$ ( since Bernoulli random variables with the same mean $\mu_X$ have the maximum variance), we also obtain,
\vspace{-0.05in}
\begin{align}
	\Pr\Big[\sum_{j = 1}^{T} X_j \geq (1+\epsilon)\mu_X T\Big] \leq \exp \big(\frac{-\epsilon^2 T \mu_X}{(2+\frac{2}{3}\epsilon)(b-a)} \big).
	\label{eq:gen_bound_2}
\end{align}
\vspace{-0.1in}

Similarly, $-Y_1, \dots, -Y_i, \dots$ also form a Martingale and applying Lemma~\ref{lem:martingale} gives the following probabilistic inequality,
\vspace{-0.05in}
\begin{align}
	\label{eq:gen_bound_1}
	\Pr\Big[\sum_{j = 1}^{T} X_j \leq (1-\epsilon)\mu_X T\Big] \leq \exp \big(-\frac{\epsilon^2 T \mu_X}{2(b-a)} \big).
\end{align}
\vspace{-0.1in}

\begin{algorithm}[!h] \small
	\caption{Robust Sampling Algorithm (\SiEIE{})}
	\label{alg:extinf}
	\KwIn{Two streams of i.i.d. random variables, $X_1, X_2, \dots$ and $X'_1, X'_2, \dots$ and $0 < \epsilon,\delta < 1$}
	\KwOut{An $(\epsilon,\delta)$-approximate $\hat \mu_X$ of $\mu_X$}
	\vspace{0.05in}
	\nonl \textbf{Step 1} \tcp{Obtain a rough estimate $\hat \mu'_X$ of $\mu_X$}
	\vspace{0.05in}
	\If{$\epsilon \geq 1/4$}{
		\textbf{return} $\hat \mu_X \leftarrow $ \GSRA{}($<X_1, X_2, \ldots>, \epsilon,\delta$)\\
	}
	$\hat \mu'_X \leftarrow $ \GSRA{}($<X_1, X_2, \ldots>, \sqrt{\epsilon},\delta/3$) \\
	\vspace{0.05in}
	\nonl \textbf{Step 2} \tcp{Estimate the variance $\hat \sigma^2_X$}
	\vspace{0.05in}
	$\Upsilon_2 = 2\frac{1 + \sqrt{\epsilon}}{1 - \sqrt{\epsilon}}(1 + \ln(\frac{3}{2})/\ln(\frac{2}{\delta}))\cdot \Upsilon; N_{\sigma} = \Upsilon_2\cdot \epsilon/\hat \mu'_X; \Delta = 0$\tcp*{$\Upsilon$ is defined the same as in Alg.~\ref{alg:stop}}
	\For{$i = 1:N_{\sigma}$}{
		$\Delta \leftarrow \Delta + (X'_{2i}-X'_{2i+1})^2$/2;\\
	}
	$\hat \rho_X = \max\{ \hat \sigma_X^2 = \Delta/N_{\sigma}, \epsilon \hat \mu'_X (b-a)\}$;\\
	\vspace{0.05in}
	\nonl \textbf{Step 3} \tcp{Estimate $\mu_X$}
	\vspace{0.05in}
	Set $T = \Upsilon_2\cdot \hat \rho_X/(\hat {\mu}'^2_X (b-a)), S \leftarrow 0$;\\
	\For{$i = 1:T$}{
		$S \leftarrow S + X_i$;
	}
	\textbf{return} $\hat \mu_X = S/T  $;\\
\end{algorithm}
\vspace{-0.05in}

\subsection{Robust Sampling Algorithm}
Our previously proposed \GSRA{} algorithm may have problem in estimating means of random variables with small variances. An important tool that we rely on to prove the approximation guarantee in \GSRA{} is the Chernoff-like bound in Eq.~\ref{eq:gen_bound_2} and Eq.~\ref{eq:gen_bound_1}. However, from the inequality in Eq.~\ref{eq:eq33}, we can also derive the following stronger inequality,
\vspace{-0.05in}
\begin{align}
	&\Pr\Big[\sum_{j = 1}^{T} X_j \geq (1+\epsilon)\mu_X T\Big] \leq \exp \Big(\frac{- \epsilon^2 T^2 \mu^2_X}{\frac{2}{3}(b-a)\epsilon\mu_X T + 2 \Var[X] T} \Big) \nonumber \\
	&\qquad \text{ }\leq \exp \Big(\frac{-\epsilon^2 T \mu^2_X}{(2+\frac{2}{3})\max\{ \epsilon\mu_X (b-a), \Var[X] \}} \Big).
	\label{eq:eq35}
\end{align}
In many cases, random variables have small variances and hence $\max\{\epsilon \mu_X (b-a),\Var[X]\} = \epsilon \mu_X (b-a)$. Thus, Eq.~\ref{eq:eq35} is much stronger than Eq.~\ref{eq:gen_bound_2} and can save a factor of $\frac{1}{\epsilon}$ in terms of required observed influences translating into the sample requirement. However, both the mean and variance are not available.

To achieve a robust sampling algorithm in terms of sample complexity, we adopt and improve the $\mathcal{AA}$ algorithm in \cite{Dagum00} for general cases of $[a,b]$ random variables. The robust sampling algorithms (\SiEIE) subsequently will estimate both the mean and variance in three steps: 1) roughly estimate the mean value with larger error ($\sqrt{\epsilon}$ or a constant); 2) use the estimated mean value to compute the number of samples necessary for estimating the variance; 3) use both the estimated mean and variance to refine the required samples to estimate mean value with desired error ($\epsilon,\delta$).


Let $X_1, X_2, \dots$ and $X'_1, X'_2, \dots$ are two streams of i.i.d random variables. Our robust sampling algorithm (\SiEIE{}) is described in Alg.~\ref{alg:extinf}. It consists of three main steps:
\begin{itemize}
	\item[1)] If $\epsilon \geq 1/4$, run \GSRA{} with parameter $\epsilon,\delta$ and return the result (Line~1-2). Otherwise, assume $\epsilon < 1/4$ and use the Generalized Stopping Rule Algorithm (Alg.~\ref{alg:stop}) to obtain an rough estimate $\hat \mu'_X$ using parameters of $\epsilon' = \sqrt{\epsilon} < 1/2, \delta' = \delta/3$ (Line~3).
	\item[2)] Use the estimated $\hat \mu'_X$ in step 1 to compute the necessary number of samples, $N_\sigma$, to estimate the variance of $X_i$, $\hat \sigma_X^2$. Note that this estimation uses the second set of samples, $X'_1, X'_2, \dots$
	\item[3)] Use both $\hat \mu'_X$ in step 1 and $\hat \sigma_X^2$ in step 2 to compute the actual necessary number of samples, $T$, to approximate the mean $\mu_X$. Note that this uses the same set of samples $X_1, X_2, \dots$ as in the first step.
\end{itemize}
The numbers of samples used in the first two steps are always less than a constant times $\Upsilon \cdot \epsilon/\mu_X$ which is the minimum samples that we can achieve using the variance. This is because the first takes the error parameter $\sqrt{\epsilon}$ which is higher than $\epsilon$ and the second step uses $N_{\sigma} = \Upsilon_2\cdot \epsilon/\hat \mu'_X$ samples.

At the end, the algorithm returns the influence estimate $\hat \mu_X$ which is the average over $T$ samples, $\hat \mu_X = S / T$.
The estimation guarantees are stated in the following theorem.

\begin{theorem}
	\label{th:extinf}
	Let $X$ be the probability distribution that $X_1, X_2, \dots$ and $X'_1, X'_2, \dots$ are drawn from. Let $\hat \mu_X$ be the estimate of $\E[X]$ returned by Alg.~\ref{alg:extinf} and $T$ be the number of drawn samples in Alg.~\ref{alg:extinf} w.r.t. $\epsilon,\delta$. We have,
\vspace{-0.05in}
	\begin{itemize}
		\item[(1)] $\Pr[\mu_X(1-\epsilon) \leq \hat \mu_X \leq (1+\epsilon)\mu_X] \geq 1-\delta$,
		\item[(2)] There is a universal constant $c'$ such that
		\vspace{-0.05in}
		\begin{align}
			\Pr[T > c' \Upsilon \rho_X/(\mu^2_X (b-a))] \leq \delta
		\end{align}
		where $\rho_Z = \max\{\epsilon \mu_X (b-a),\Var[X]\}$.
	\end{itemize}
\vspace{-0.1in}
\end{theorem}

Compared to the $\mathcal{AA}$ algorithm in \cite{Dagum00}, first of all, we replace their stopping rule algorithm with \GSRA{} and also, we change the computation of $\Upsilon_2$ which is always smaller than that of \cite{Dagum00} by a factor of $1+\sqrt{\epsilon}-2\epsilon \geq 1$ when $\epsilon \leq 1/4$.

%% file: body/infest.tex
\vspace{-0.05in}
\section{Influence Estimation at Scale}
\label{sec:inf}

This section applies our \SiEIE{} algorithm to estimate both the outward influence and the traditional influence spread.

\vspace{-0.1in}
\subsection{Outward Influence Estimation}
We directly apply \SiEIE{} algorithm on two streams of i.i.d. random variables $\Ys_1, \Ys_2, \dots$ and $\pYs_1, \pYs_2, \dots$, which are generated by \IICP{} sampling algorithm, with the precision parameters $\epsilon,\delta$.

The algorithm is called \textit{Scalable Outward Influence Estimation Algorithm} (\OInf{}) and presented in Alg.~\ref{alg:soiea} which generates two streams of random variables $\Ys_1, \Ys_2, \dots$ and $\pYs_1, \pYs_2, \dots$ (Line~1) and applies \SiEIE{} algorithm on these two streams (Line~2). Note that outward influence estimate is achieved by scaling down $\mu_Y$ by $\beta_0$ (Lemma~\ref{lem:sim_prob}).
\begin{algorithm} \small
	\caption{\OInf{} Alg. to estimate outward influence}
	\label{alg:soiea}
	\KwIn{A probabilistic graph $\G$, a set $S$ and $\epsilon,\delta$}
	\KwOut{$\hat \I(S)$ - an $(\epsilon,\delta)$-estimate of $\I(S)$}
	Generate two streams of i.i.d. random variables $\Ys_1, \Ys_2, \dots$ and $\pYs_1, \pYs_2, \dots$ by \IICP{} algorithm. \\
	\textbf{return} $\hExt(S) \leftarrow \beta_0\cdot \SiEIE(<\Ys_1, \dots>, <\pYs_1, \dots>, \epsilon,\delta)$
\end{algorithm}

We obtain the following theoretical results incorporated from Theorem~\ref{th:extinf} of \SiEIE{} and \IICP{} samples.
\begin{theorem}
	The \OInf{} algorithm gives an $(\epsilon,\delta)$ outward influence estimation. The observed outward influences (sum of $\Ys{}$) and the number of generated random variables are in $O(\ln(\frac{2}{\delta})\frac{1}{\epsilon^2}\frac{\rho_Y}{\Ext(S)/\beta_0})$ and  $O(\ln(\frac{2}{\delta})\frac{1}{\epsilon^2}\frac{\rho_Y}{\Ext^2(S)/\beta_0^2})$ respectively, where $\rho_Y = \max\{\epsilon\Ext(S)(n-|S|-1)/\beta_0, \Var[\Ys_i]\}$.
\end{theorem}

Note that $\E[\Ys{}] = \Ext(S)/\beta_0 \geq 1$.

\vspace{-0.05in}
\subsection{Influence Spread Estimation}

Not only is the concept of \textit{outward influence} helpful in discriminating the relative influence of nodes  but also its sampling technique, \IICP, can help scale up the estimation of influence spread (\IE{}) to billion-scale networks.


\vspace{0.05in}
\textbf{Naive approach}. A naive approach is to 1) obtain an $(\epsilon, \delta)$-approximation $\hat \I_{out}(S)$ of $\Ext(S)$ using Monte-Carlo estimation 2) return $\hExt(S) + |S|$. 
It is easy to show that this approach return an $(\epsilon, \delta)$-approximation for $\I(S)$. 
This approach will require $O(\ln(\frac{2}{\delta}) \frac{1}{\epsilon^2} n)$ \IICP{} random samples.


However, the naive approach is not optimized to estimate influence due to several reasons: 1) a loose bound $\mu_Y = \E[\Ys{}] \geq 1$ is applied to estimate outward influence; 2) casting from $(\epsilon,\delta)$-approximation of outward influence to $(\epsilon,\delta)$-approximation of influence introduces a gap that can be used to improve the estimation guarantees. We next propose more efficient algorithms based on Importance IC Sampling to achieve an $(\epsilon,\delta)$-approximate of both outward influence and influence spread. Our methods are based on two effective mean estimation algorithms.

\textbf{Our approach}. Based on the observations that 
	\begin{itemize}
		\item $1 \leq Y^{(S)} \leq n-|S|$, i.e., we know better bounds for $Y^{(S)}$ in comparison to the cascade size which is in the range $[1, n]$.
		\item As we want to have an $(\epsilon, \delta)$-approximation for $Y^{(S)} + |S|$, the fixed add-on $|S|$ can be leveraged to reduce the number of samples.
	\end{itemize}
	


We combine the effective \SiEIE{} algorithm with our Importance IC Polling (\IICP{}) for estimating the influence spread of a set $S$. For influence spread estimation, we will analyze random variables based on samples generated by our Importance IC Polling scheme and use those to devise an influence estimation algorithm.

Since outward influence and influence spread differ by an additive factor of $|S|$, for each outward sample $\Ys{}$ generated by \IICP{}, let define a corresponding variable $\Zs{}$,
\vspace{-0.05in}
\begin{align}
	\label{eq:z_s}
	\Zs{} = \Ys{}\cdot \beta_0 + |S|,
\end{align} 
\vspace{-0.15in}

\noindent where $\beta_0$ is defined in Eq.~\ref{eq:zerop}. We obtain,
\vspace{-0.05in}
\begin{itemize}
	\item $|S| + \beta_0 \leq \Zs{} \leq |S| + \beta_0 (n-|S|)$,
	\item $\E[\Zs{}] = \E[\Ys{}]\cdot \beta_0 + |S| = \Ext(S) + |S| = \I(S)$,
\end{itemize}
\vspace{-0.05in}
and thus we can to approximate $\I(S)$ by estimating $\E[\Zs{}]$.

Recall that to estimate the influence $\I(S)$ of a seed set $S$, all the previous works \cite{Kempe03,Leskovec07,Chen10} resort to simulating many influence cascades from $S$ and take the average size of those generated cascades. Let call $\Ms{}$ the random variable representing the size of such a influence cascade. Then, we have $\E[\Ms{}] = \I(S)$. Although both $\Zs{}$ and $\Ms{}$ can be used to estimate the influence, they have different variances that lead to difference in convergence speed when estimating their means. The relation between variances of $\Zs{}$ and $\Ms{}$ is stated as follows.
\begin{Lemma}
	\label{lem:var}
	Let $\Zs{}$ defined in Eq.~\ref{eq:z_s} and $\Ms{}$ be random variable for the size of a influence cascade, the variances of $\Zs{}$ and $\Ms{}$ satisfy,
	\vspace{-0.1in}
	\begin{align}
		\Var[\Zs{}] = \beta_0 \cdot \Var[M^{(S)}] - (1-\beta_0)\Ext^2(S)
	\end{align}
\end{Lemma}


Note that $0 \leq \beta_0 \leq 1$ and $\I(S) \geq |S|$. Thus, the variance of $\Zs{}$ is much smaller than $\Ms{}$. Our proposed \SiEIE{} on random variables $X_i$ makes use of the variances of random variables and thus, benefits from the small variance of $\Zs{}$ compared to the same algorithm on the previously known random variables $\Ms{}$.

\begin{algorithm} \small
	\caption{\OBInf{} Alg. to estimate influence spread}
	\label{alg:obiea}
	\KwIn{A probabilistic graph $\G$, a set $S$ and $\epsilon,\delta$}
	\KwOut{$\hat \I(S)$ - an $(\epsilon,\delta)$-estimate of $\I(S)$}
	Generate two streams of i.i.d. random variables $\Ys_1, \Ys_2, \dots$ and $\pYs_1, \pYs_2, \dots$ by \IICP{} algorithm. \\
	Compose two streams $\Zs_1, \Zs_2, \dots$ and $\pZs_1, \pZs_2, \dots$ from $\Ys_1, \Ys_2, \dots$ and $\pYs_1, \pYs_2, \dots$ using Eq.~\ref{eq:z_s}.\\
	\textbf{return} $\hat \I(S) \leftarrow $\SiEIE($<\Zs_1, \dots>, <\pZs_1, \dots>, \epsilon,\delta$)
\end{algorithm}

Thus, we apply the \SiEIE{} on random variables generated by \IICP{} to develop Scalable Influence Estimation Algorithm (\OBInf{}). \OBInf{} is described in Alg.~\ref{alg:obiea} which consists of two main steps: 1) generate i.i.d. random variables by \IICP{} and 2) convert those variables to be used in \SiEIE{} to estimate influence of $S$. The results are stated as follows,

%

%

\begin{theorem}
	The \OBInf{} algorithm gives an $(\epsilon,\delta)$ influence spread estimation. The observed influences (sum of random variables $\Zs{}$) and the number of generated random variables are in $O(\ln(\frac{2}{\delta})\frac{1}{\epsilon^2}\frac{\rho_Z}{\I(S)})$ and  $O(\ln(\frac{2}{\delta})\frac{1}{\epsilon^2}\frac{\rho_Z}{\I^2(S)})$, where $\rho_Z = \max\{\epsilon\I(S)\beta_0(n-|S|-1), \Var[\Zs_i]\}$.
\end{theorem}

\textbf{Comparison to \KDD{} \cite{Lucier15}.} Compared to the most recent state-of-the-art influence estimation in \cite{Lucier15} that requires $O(\frac{n \log^5(n)}{\epsilon^2})$ observed influences, the \OBInf{} algorithm incorporating \IICP{} sampling with \textsf{RSA} saves at least a factor of $\log^4(n)$. That is because the necessary observed influences in \OBInf{} is bounded by $O(\ln(\frac{2}{\delta})\frac{1}{\epsilon^2}\frac{\beta_0 \rho_Z}{\I(S)})$. Since $\Var[\Zs_i] \leq \I(S)(|S| + \beta_0 (n-|S|)-\I(S)) \leq \I(S) (n-|S|-1)$ and hence, $\rho_Z \leq \I(S) (n-|S|-1)$, when $\delta = \frac{1}{n}$ as in \cite{Lucier15}, the observed influences is then,
\vspace{-0.05in}
\begin{align}
	O(\ln(\frac{2}{\delta})\frac{1}{\epsilon^2}\frac{\rho_Z}{\I(S)}) \leq O(\frac{n \log(2/\delta)}{\epsilon^2}) \leq O(\frac{n \log(n)}{\epsilon^2})
\end{align}
\vspace{-0.15in}

\noindent Consider $\epsilon,\delta$ as constants, the observed influences is $O(n)$.

\vspace{-0.1in}
\subsection{Influence Spread  under other Models}
\label{subsec:extension}
We can easily apply the \SiEIE{} estimation algorithm to obtain an $(\epsilon,\delta)$-estimate of the influence spread under other cascade models as long as there is a Monte-Carlo sampling procedure to generate sizes of random cascades. For most stochastic diffusion models, including both discrete-time models, e.g. the popular LT with a naive sample generator described in \cite{Kempe03}, SI and SIR \cite{Daley01} or their variants with deadlines \cite{Nguyen165}, and continuous-time models \cite{Du13}, designing such a Monte-Carlo sampling procedure is straightforward. Since the influence cascade sizes are at least the seed size, we always needs at most $O(n)$ samples.

To obtain  more efficient sampling procedures, we can extend the idea of sampling non-trivial cascade in \IICP{} to other models. Such sampling procedures in general will result in random variables with smaller variances and tighter bounds on the ranges. In turns,  \SiEIE{}, that benefits from smaller variance and range, will requires fewer samples for estimation.


\vspace{-0.1in}
\subsection{Parallel Estimation Algorithms}

We develop the parallel versions of our algorithms to speed up the computation and demonstrate the easy-to-parallelize property of our methods. Our main idea is that the random variable generation by \IICP{} can be run in parallel. In particular, random variables used in each step of the core \SiEIE{} algorithm can be generated simultaneously. 
Recall that \IICP{} only needs to store a queue of newly active nodes, an array to mark the active nodes and a single variable $Y^{(S)}$. In total, each thread requires space in order of the number of active nodes in that simulation, $O(Y^{(S)})$, which is at most linear with size of the graph $O(n)$. In fact due to the stopping condition of linear number of observed influences, the total size of all the threads is bounded by $O(n)$ assumed the number of threads is relatively small compared to $n$.

Moreover, our algorithms can be implemented efficiently in terms of communication cost in distributed environments. This is because the output of \IICP{} algorithm is just a single number $Y^{(S)}$ and thus, worker nodes in a distributed environment only communicate that single number back to the node running the estimation task. Here each \IICP{} node holds a copy of the graph. However, the programming model needs to be considered carefully. For instance, as pointed out in many studies that the famous MapReduce is not a good fit for iterative graph processing algorithms \cite{Gupta13,Lin10}. 

%% file: body/experiment.tex
\vspace{-0.05in}
\section{Experiments}
\label{sec:exp}

We will experimentally show that Outward Influence Estimation (\OInf{}) and Outward-Based Influence Estimation (\OBInf{}) are not only several orders of magnitudes faster than existing state-of-the-art methods but also consistently return much smaller errors. We present empirical validation of our methods on both real world and synthetic networks.

\vspace{-0.1in}
\subsection{Experimental Settings}
\textbf{Algorithms.} We compare performance of \OInf{} and \OBInf{} with the following algorithms:
\begin{itemize}
	\item \KDD{} \cite{Lucier15}: A recent influence estimation algorithm by Lucier et al. \cite{Lucier15} in KDD'15 that provides approximation guarantees. We reimplement the algorithm in C++ accordingly to the description in \cite{Lucier15}\footnote{Through communication with the authors of \cite{Lucier15}, the released code has some problem and is not ready for testing.}.
	\item \MCk{}, \MCb{}: Variants of Monte-Carlo method that generates the traditional influence cascades \cite{Kempe03,Leskovec07} to estimate (outward) influence spread.
	\item \MCe{}: The Monte-Carlo method that uses the traditional influence cascades and guarantees $(\epsilon,\delta)$-estimation. Following \cite{Lucier15}, \MCe{} is only for measuring the running time of the normal Monte-Carlo to provide the same $(\epsilon,\delta)$-approximation guarantee. In particular, we obtain running time of \MCe{} by interpolating from that from \MCk{}, i.e. $\frac{1}{\epsilon^2}\ln(\frac{1}{\delta})n\frac{\textsf{Time}(\text{\MCk{}})}{10000}$.
\end{itemize}


\renewcommand{\arraystretch}{1}
\setlength\tabcolsep{3pt}
\begin{table}[!h]
	\vspace{-0.08in}
	\caption{Datasets' Statistics}
	\vspace{-0.15in}
	\label{tab:data_sum}
	\centering
	\begin{tabular}{ l  r  r  r}\toprule
		\textbf{Dataset} & \bf \#Nodes& \bf \#Edges & \bf Avg. Degree\\
		\midrule
		NetHEP\footnote{From http://snap.stanford.edu}& 15K & 59K & 4.1\\
		NetPHY\footnotemark[\value{footnote}] & 37K & 181K & 13.4\\
		Epinions\footnotemark[\value{footnote}] & 75K & 841K & 13.4\\
		DBLP\footnotemark[\value{footnote}] & 655K & 2M & 6.1\\
		Orkut\footnotemark[\value{footnote}] & 3M & 117M & 78.0\\
		Twitter \cite{Kwak10} & 41.7M & 1.5G & 70.5\\
		Friendster\footnotemark[\value{footnote}] & 65.6M & 1.8G & 54.8\\
		\bottomrule
		\multicolumn{4}{r}{\footnotemark[\value{footnote}]From http://snap.stanford.edu}\\
	\end{tabular}
\end{table}
\renewcommand{\arraystretch}{0.9}
\setlength\tabcolsep{5pt}
\begin{table*}[!ht] \small
	\centering
	\caption{Comparing performance of algorithms in estimating outward influences}
	\vspace{-0.1in}	
	\begin{tabular}{clrrrrrrrrrrrr}
		\toprule
		\multirow{1}{*}{} & \multirow{1}{*}{} & \multicolumn{3}{c}{\textbf{Avg. Rel. Error (\%)}} && \multicolumn{3}{c}{\textbf{Max. Rel. Error (\%)}} && \multicolumn{4}{c}{\textbf{Running time (sec)}}  \\
		\cmidrule{3-5} \cmidrule{7-9} \cmidrule{11-14}
		
		\textbf{Dataset} & \textbf{Edge Models} & \OInf{} & \MCk{} & \MCb{} && \OInf{} & \MCk{} & \MCb{} && \OInf{} & \MCk{} & \MCb{} & \MCe{} \\
		\midrule	
		
		\multirow{4}{*}{NetHEP} & $wc$ & 0.3 & 1.9 & 0.6 && 2.3 & 25.0 & 8.9 && 0.1 & 0.1 & 0.1 & 12.3 \\ 
		& $p=0.1$ & 1.0 & 3.7 & 1.2 && 9.7 & 63.0 & 17.2 && 0.2 & 0.1 & 1.0 & 149.5 \\ 
		& $p=0.01$ & 0.0 & 4.5 & 1.6 && 0.2 & 20.2 & 9.2 && 0.2 & 0.1 & 0.1 & 8.8 \\ 
		& $p=0.001$ & 0.0 & 19.2 & 4.6 && 0.1 & 100.0 & 26.4 && 0.2 & 0.1 & 0.1 & 8.5 \\ 
		\midrule
		\multirow{4}{*}{NetPHY} & $wc$ & 0.1 & 1.4 & 0.4 && 1.5 & 32.8 & 6.2 && 0.4 & 0.1 & 0.1 & 34.7 \\ 
		& $p=0.1$  & 0.5 & 4.0 & 1.3 && 6.6 & 46.3 & 18.5 && 0.5 & 0.1 & 0.5 & 203.0 \\ 
		& $p=0.01$ & 0.0 & 5.5 & 1.7 && 0.2 & 30.4 & 10.7 && 0.6 & 0.1 & 0.1 & 25.0 \\ 
		& $p=0.001$ & 0.0 & 19.1 & 5.1 && 0.0 & 80.0 & 28.1 && 0.7 & 0.1 & 0.1 & 24.0 \\ 
		\bottomrule	
	\end{tabular}
	\label{tbl:set1}
	\vspace{-0.35in}
\end{table*}

\begin{figure*}[!t]
	\centering
	\subfloat[WC model]{
		\includegraphics[width=0.24\linewidth]{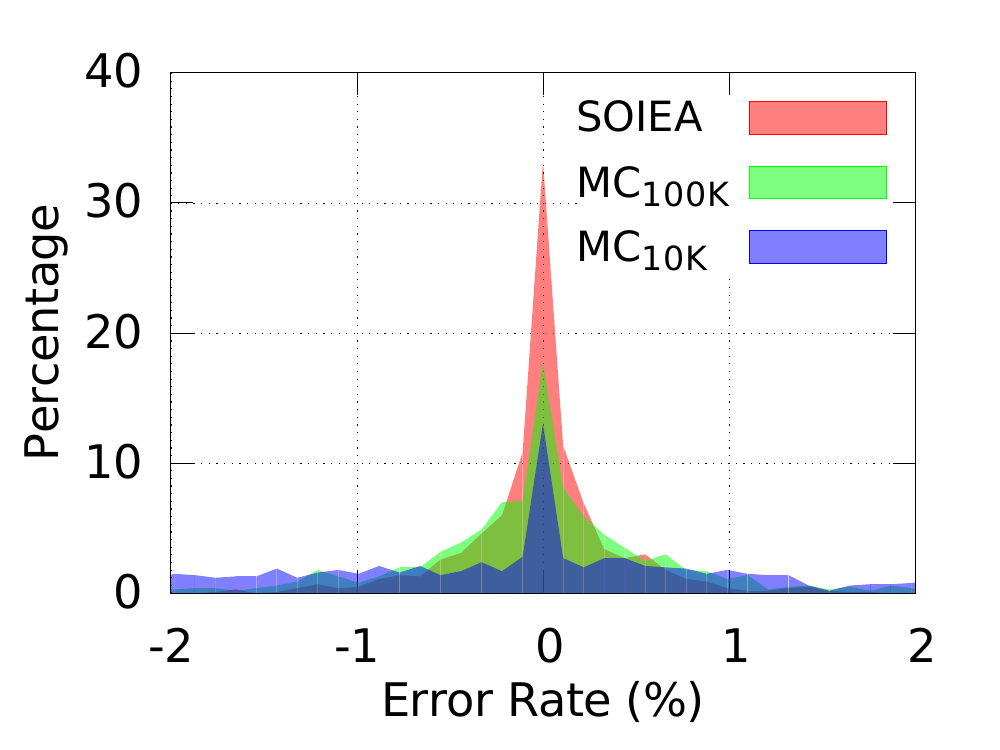}
	}
	\subfloat[$p=0.1$]{
		\includegraphics[width=0.24\linewidth]{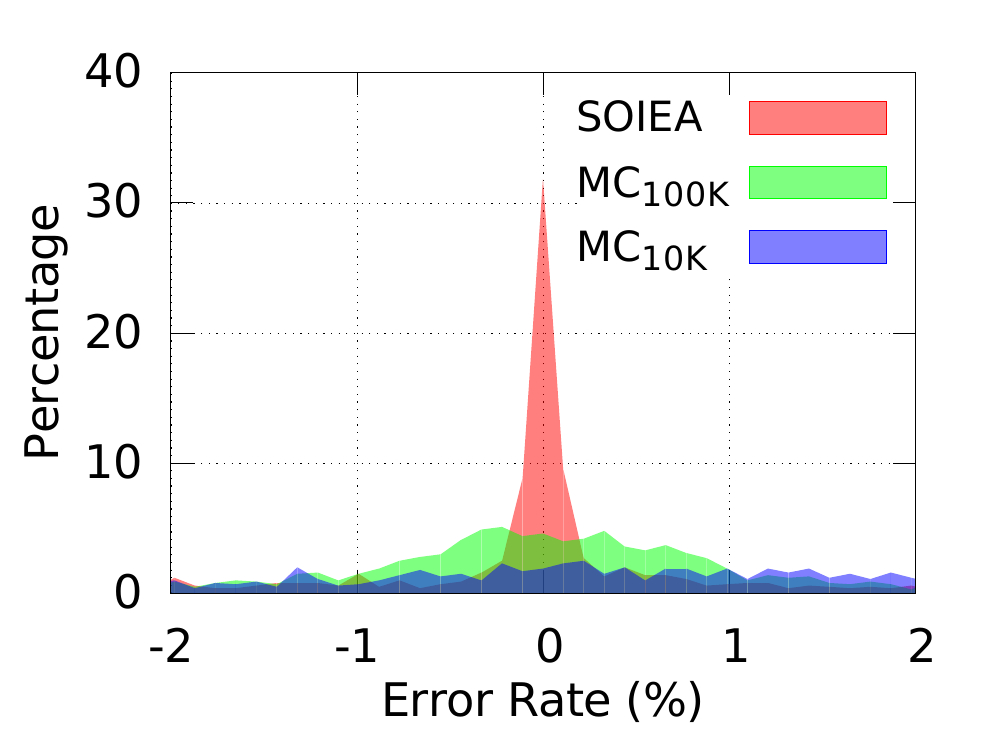}
	}
	\subfloat[$p=0.01$]{
		\includegraphics[width=0.24\linewidth]{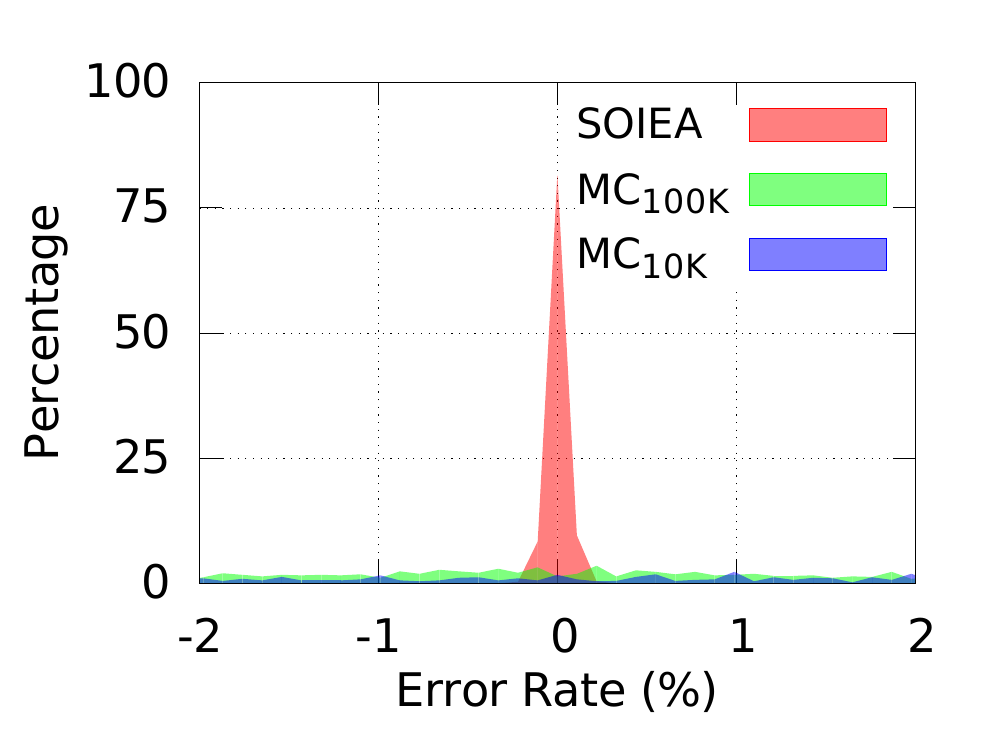}
	}
	\subfloat[$p=0.001$]{
		\includegraphics[width=0.24\linewidth]{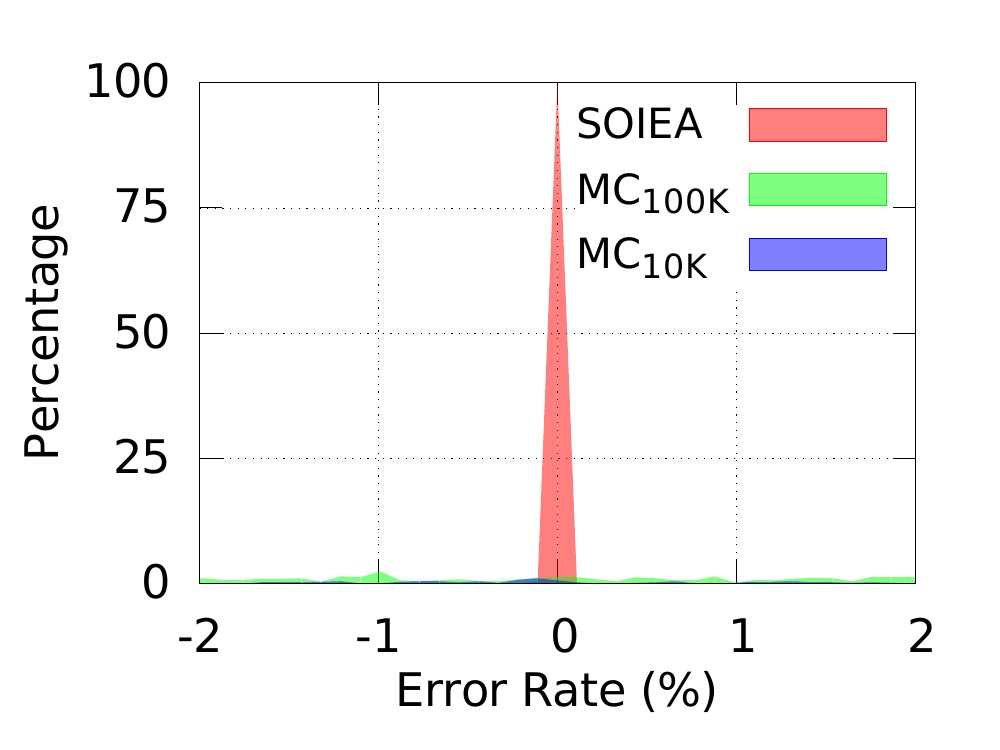}
	}
	\vspace{-0.15in}
	\caption{Error distributions (histogram) of the approximation errors of \OInf{}, \MCk{}, \MCb{} on NetHEP}
	\label{fig:set1}
	\vspace{-0.2in}
\end{figure*}

%

\textbf{Datasets.} 
We use both real-world networks and synthetic networks generated by GTgraph \cite{Bader06}. For real world networks, we choose a set of 7 datasets with sizes from tens of thousands to 65.6 millions. Table~\ref{tab:data_sum} gives a summary. GTgraph generates synthetic graphs with varying number of nodes and edges.





\textbf{Metrics.}
We compare the performance of the algorithms in terms of solution quality and running time. To compare the solution quality, we adopt the relative error which  shows how far the estimated number from the ``ground truth". The relative error of outward influence is computed as follows:
\vspace{-0.05in}
\begin{align}
	|\frac{\hExt(S)}{\Ext(S)} - 1| \cdot 100\%
\end{align}
\vspace{-0.1in}

\noindent where $\hExt(S)$ is estimated outward influence of seed set $S$ by the algorithm, $\Ext(S)$ is ``ground truth" for $S$.

Similarly, relative error of influence spread is,
\vspace{-0.05in}
\begin{align}
|\frac{\hat \I(S)}{\I(S)} - 1| \cdot 100\%
\end{align}
\vspace{-0.15in}

We test the algorithms on estimating different seed set sizes. For each size, we generate a set of 1000 random seed sets. We will report the average relative error (Avg. Rel. Error) and maximum relative error (Max. Rel. Error).

\textbf{Ground-truth computation.} We use estimates of influence and outward influence with a very small error corresponding to the setting $\epsilon = 0.005, \delta=1/n$. We note that previous researches \cite{Lucier15, Tang15} compute the ``ground truth" by running Monte-Carlo with 10,000 samples which is not sufficient as we will show later in our experiments.




\textbf{Parameter Settings.}
For each of the datasets, we consider two common edge weighting models:
\begin{itemize}
	\item \textsf{Weighted Cascade} (\textsf{WC}): The weight of edge $(u,v)$ is calculated as	$w(u,v) = \frac{1}{d_{in}(v)}$ where $d_{in}(v)$ denotes the in-degree of node $v$, as in \cite{Chen10,Tang14,Cohen14,Tang15,Nguyen163}.
	\item \textsf{Constant model}: All the edges has the same constant probability $p$ as in \cite{Kempe03, Chen10, Cohen14}. We consider three different values of $p$, i.e. $0.1, 0.01, 0.001$.
\end{itemize}

We set $\epsilon = 0.1$, $\delta = 1/n$ for \OInf{} and \OBInf{} by default or explicitly stated otherwise.

\textbf{Environment.} 
All algorithms are implemented in C++ and compiled using GCC 4.8.5. We conduct all experiments on a CentOS 7 workstation with two Intel Xeon 2.30GHz CPUs adding up to 20 physical cores and 250GB RAM. 

\vspace{-0.1in}
\subsection{Outward Influence Estimation}

We compare \OInf{} against \MCk{} and \MCb{} in four different edge models on NetHEP and NetPHY dataset. The results are presented in Table~\ref{tbl:set1} and Figure~\ref{fig:set1}.


\vspace{-0.1in}
\subsubsection{Solution Quality}

Table~\ref{tbl:set1} illustrates that the outward influences computed by \OInf{} consistently have much smaller errors in both average and maximum cases than \MCk{} and \MCb{} in all edge models. In particular, on NetHEP with $p=0.001$ edge model, \OInf{} has average relative error close to 0\% while it is 19.2\% and 4.6\% for \MCk{}, \MCb{} respectively; the maximum relative errors of \MCk{}, \MCb{} in this case are 100\%, 26.4\% which are much higher than \OInf{} of 0.1\%. Apparently, \MCb{} has smaller error rate than \MCk{} since it uses 10 times more samples.

Figure \ref{fig:set1} shows error distributions of \OInf{}, \MCk{}, and \MCb{} on NetHEP. In all considered edge models, \OInf{}'s error highly concentrates around 0\% while errors of \MCk{} and \MCb{} wildly spread out to a very large spectrum. In particular, \OInf{} has a huge spike at the 0 error while both \MCk{} and \MCb{} contain two heavy tails in two sides of their error distributions. Moreover, when $p$ gets smaller, the tails get larger as more and more empty influence simulations are generated in the traditional method.

\subsubsection{Running Time}


From Table~\ref{tbl:set1}, the running time of \MCk{} and \MCb{} is close to that of \OInf{} while \MCe{} takes up to 700 times slower than the others. Thus, in order to achieve the same approximation guarantee as \OInf{}, the naive Monte-Carlo will need 700 more time than \OInf{}.

Overall, \OInf{} achieves significantly better solution quality and runs substantially faster than Monte-Carlo method. With larger number of samples, Monte-Carlo method can improve the quality but the running time severely suffers.

\vspace{-0.1in}
\subsection{Influence Spread Estimation}

This experiment evaluates \OBInf{} by comparing its performance with the most recent state-of-the-art \KDD{} and naive Monte-Carlo influence estimation. Here, we use \textsf{WC model} to assign probabilities for the edges. We set the $\epsilon$ parameter for \KDD{} to 0.4 since we cannot run with smaller value of $\epsilon$ for this algorithm. Note that \KDD{} guarantees an error of $(1+8\epsilon)$, \textit{which is equivalent to a maximum relative error of 320\%.} For a fair comparison, we also run \OBInf{} with $\epsilon=0.4$. We use the gold-standard 10000 samples for the Monte-Carlo method (\MCk{}).  We set a time limit of 6 hours for all algorithms.

\setlength\tabcolsep{7pt}
\begin{table*}[!t] \small
	\centering
	\caption{ Comparing performance of algorithms in estimating influence spread in \textsf{WC Model} (seed set size $|S|=1$)}
	\vspace{-0.15in}
	\begin{tabular}{lrrrrrrrrrrrrr}
		\toprule
		\multirow{2}{*}{} & \multicolumn{3}{c}{\textbf{Avg. Rel. Error (\%)}} && \multicolumn{3}{c}{\textbf{Max. Rel. Error (\%)}} && \multicolumn{5}{c}{\textbf{Running time (sec.)}}  \\
		\cmidrule{2-4} \cmidrule{6-8} \cmidrule{10-14}
		\textbf{Dataset} & \OBInf{} & \MCk{} & \KDD{} && \OBInf{} & \MCk{} & \KDD{} && \OBInf{} & \OBInf{} \textsf{(16 cores)} & \MCk{} & \MCe{} & \KDD{} \\		
		\midrule
		NetHEP & 0.2 & 1.2 & 17.7 && 1.5 & 6.6 & 82.7 && 0.1 & 0.1 & 0.0 & 0.8 & 3417.6 \\ 
		NetPHY & 0.1 & 0.4 & 22.9 && 0.6 & 5.3 & 43.0 && 0.1 & 0.1 & 0.0 & 2.6 & 8517.7 \\ 
		Epinions & 0.9 & 5.3 & n/a && 5.2 & 19.7 & n/a && 0.2 & 0.1 & 0.0 & 21.9 & n/a \\ 
		DBLP & 0.3 & 1.2 & n/a && 1.9 & 8.7 & n/a && 2.8 & 1.3 & 0.1 & 770.4 & n/a \\ 
		Orkut & 0.5 & 3.0 & n/a && 3.2 & 16.0 & n/a && 54.2 & 4.76 & 2.9 & $8.2\cdot 10^4$ & n/a \\ 
		Twitter & 1.0 & 37.1 & n/a && 3.1 & \textbf{240.8} & n/a && 1272.3 & 106.2 & 7.9 & $3.5\cdot 10^6$ & n/a \\ 
		Friendster & 0.1 & 3.1 & n/a && 0.6 & 23.6 & n/a && 1510.1 & \textbf{165.1} & 2.8 & $2.1\cdot 10^6$ & n/a \\	
		\bottomrule		
	\end{tabular}
	\label{tbl:set2a}
	\vspace{-0.1in}
\end{table*}

%
\setlength\tabcolsep{6pt}
\begin{table*}[!t] \small
	\centering
	\caption{ Comparing performance of algorithms in estimating influence spread in \textsf{WC Model} (seed set size $|S|=5\%|V|$)}
	\vspace{-0.15in}
	\begin{tabular}{lrrrrrrrrrrrrr}
		\toprule
		\multirow{2}{*}{} & \multicolumn{3}{c}{\textbf{Avg. Rel. Error (\%)}} && \multicolumn{3}{c}{\textbf{Max. Rel. Error (\%)}} && \multicolumn{5}{c}{\textbf{Running time (sec.)}}  \\
		\cmidrule{2-4} \cmidrule{6-8} \cmidrule{10-14}
		\textbf{Dataset} & \OBInf{} & \MCk{} & \KDD{} && \OBInf{} & \MCk{} & \KDD{} && \OBInf{} & \OBInf{} (16 cores) & \MCk{} & \MCe{} & \KDD{} \\
		\midrule						
		NetHEP & 0.1 & 0.0 & 11.1 && 0.4 & 0.2 & 14.1 && 0.1 & 0.1 & 2.1 & 191.7 & 600.5  \\ 
		NetPHY & 0.1 & 0.0 & 24.4 && 0.2 & 0.1 & 26.3  && 0.1 & 0.1 & 5.3 & 1297.1 & 3326.4  \\ 
		Epinions & 0.2 & 0.1 & 20.2 && 0.4 & 0.2 & 23.8 && 0.3 & 0.1 & 20.1 & $1.1\cdot 10^4$ & 9325.6  \\ 	 
		DBLP  & 0.0 & 1.8 & n/a && 0.2 & 1.9 & n/a && 3.5 & 0.3 & 184.9 & $1.0\cdot 10^6$ & n/a  \\ 
		Orkut & 0.1 & 0.0 & n/a && 0.7 & 0.1 & n/a && 51.6 & 4.6  & 5322.8 & $1.5\cdot 10^8$ & n/a  \\ 
		Twitter & 0.2 & n/a & n/a && 0.5 & n/a & n/a && 1061.6 & 93.5 & n/a & n/a & n/a  \\ 
		Friendster & 0.1 & n/a & n/a && 0.2 & n/a & n/a && 2068.8 & \textbf{183.1} & n/a & n/a & n/a  \\ 		
		\bottomrule
	\end{tabular}
	\label{tbl:set2b} 
	\vspace{-0.15in}
\end{table*}

\subsubsection{Solution Quality}

Table~\ref{tbl:set2a} presents the solution quality of the algorithms in estimating size 1 seed sets, i.e. $|S|=1$. It shows that \OBInf{} consistently achieves substantially higher quality solution than both \KDD{} and \MCk{}. Note that \KDD{} can only run on NetHEP and NetPHY under time limit. The average relative error of \KDD{} is 88 to 229 times higher than \OBInf{} while its maximum relative error is up to 82\% compared to the ground truth. The large relative error of \KDD{} is explained by its loose guaranteed relative error (320\%). Whereas, the average relative error of \MCk{} is up to 37 times higher than \OBInf{}. The maximum relative error of \MCk{} is up to 240\% higher than the ground truth on Twitter dataset that demonstrates the insufficiency of using 10000 traditional influence samples to get the ground truth.

Differ from Table~\ref{tbl:set2a}, Table~\ref{tbl:set2b} shows the results in estimating influences of seed sets of size $5\%$ the total number of nodes. Under 6 hour limit, \KDD{} can only run on NetHEP, NetPHY, and Epinions while \MCk{} could not handle the large Twitter and Friendster graph. \KDD{} still has a very high error compared to the other two while \OBInf{} and \MCk{} returns the similar quality solutions. This is because $5\%$ of the nodes is an enormous number, i.e. $>1000000$ for Friendster, and thus, the influence is huge and very few samples are needed regardless of using the traditional method or \IICP{}.

\subsubsection{Running Time}

In both cases of two seed set sizes, \OBInf{} vastly outperforms \MCe{} and \KDD{} by several orders of magnitudes. \KDD{} is up to $10^5$ times slower than \OBInf{} and can only run on small networks, i.e. NetHEP, NetPHY and Epinions. Compared with \MCe{}, the speedup factor is around $10^4$, thus, \MCk{} cannot run for the two largest networks, Twitter and Friendster in case $|S| = 5\%|V|$.

We also test the parallel version of \OBInf{}. With 16 cores, \OBInf{} runs about 12 times faster than that on a single core in large networks achieving an effective factor of around 75\%.

%

Overall, \OBInf{} consistently achieves much better solution quality and run significantly fastest than \KDD{} and the naive \MC{} method. Surprisingly, under time limit of 6 hours, \KDD{} can only handle small networks and has very high error. The \MC{} method achieves better accuracy for large seed sets, however, its running time increases dramatically resulting in failing to run on large datasets.

\begin{figure}[!ht]
	\vspace{-0.13in}
	\centering
	\subfloat[Running time in linear]{
		\includegraphics[width=0.48\linewidth]{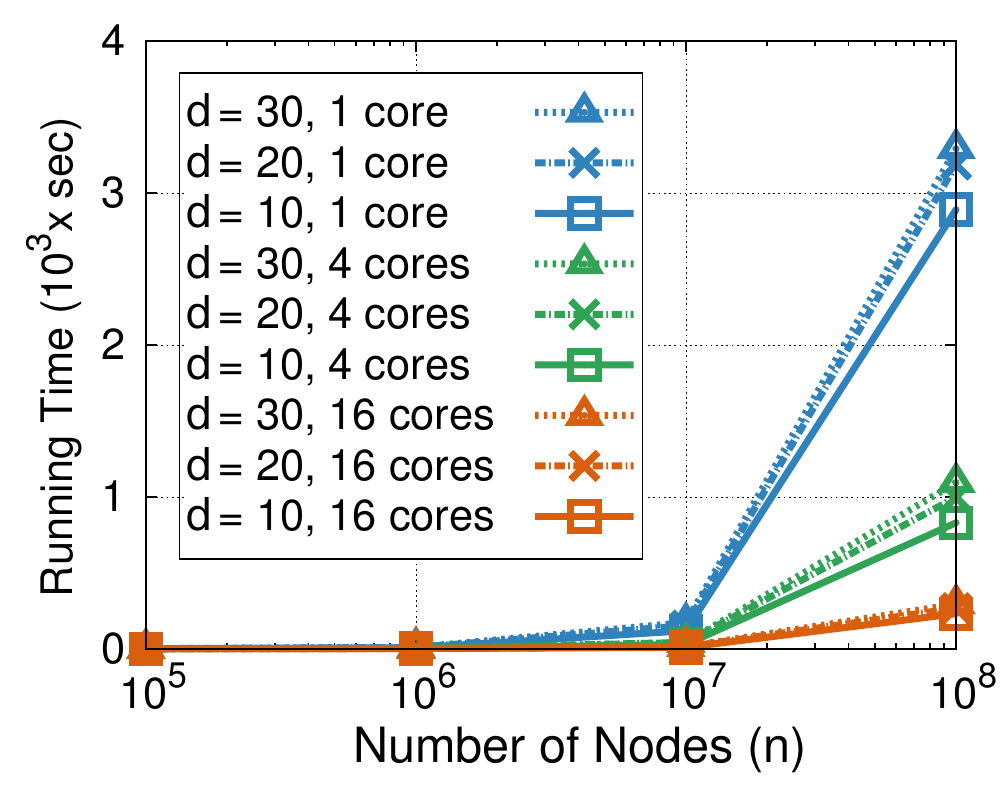}
	}
	\subfloat[Running time in log scale]{
		\includegraphics[width=0.48\linewidth]{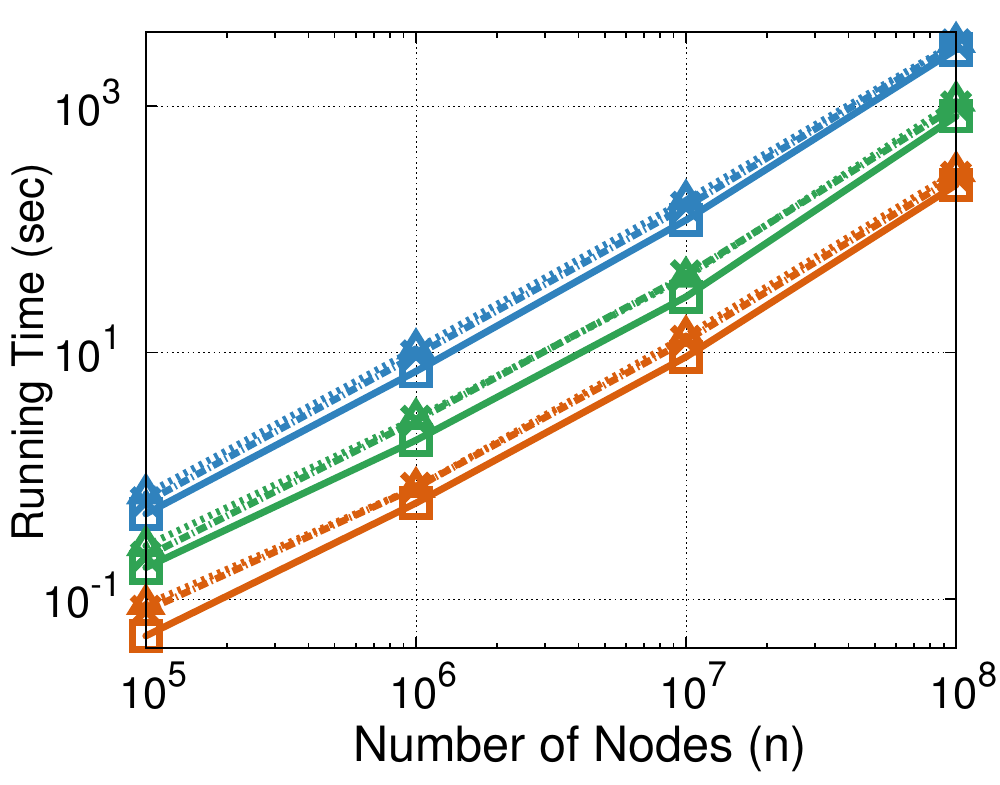}
	}
			
	\vspace{-0.12in}
	\caption{Running time of \OBInf{} on synthetic networks}
	\label{fig:set2_syn}
	\vspace{-0.15in}
\end{figure}

\begin{figure}[!t]
	\vspace{-0.15in}
	\centering
	\subfloat[Relative Error]{
		\includegraphics[width=0.49\linewidth]{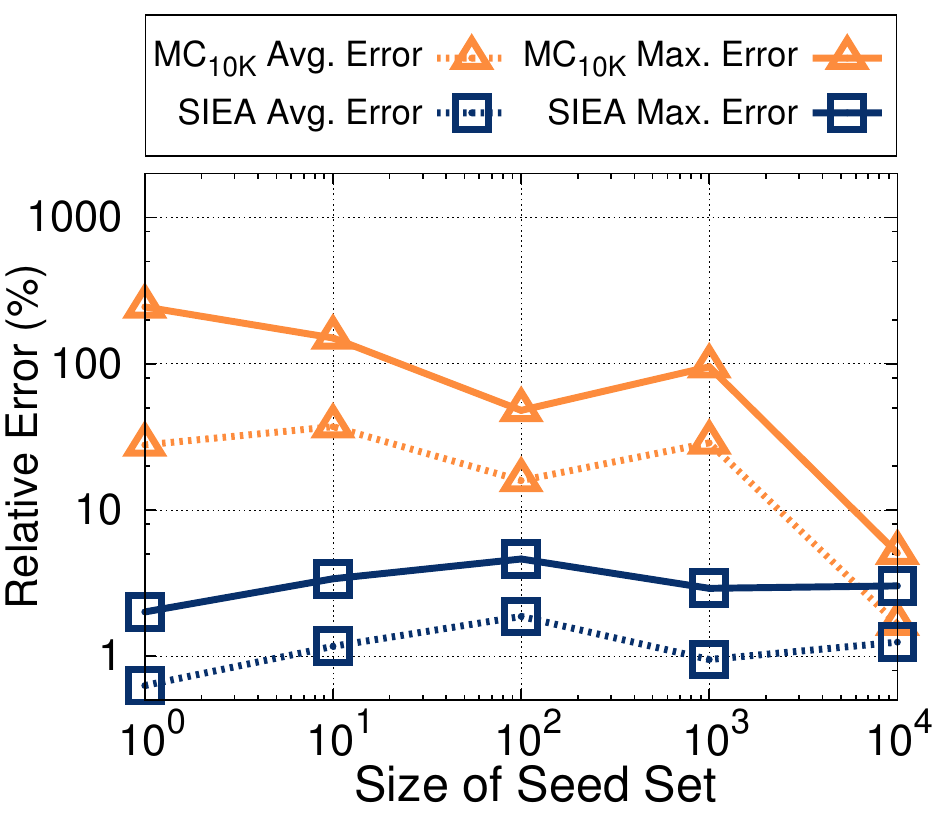}
	}
	\subfloat[Running Time]{
	\includegraphics[width=0.47\linewidth]{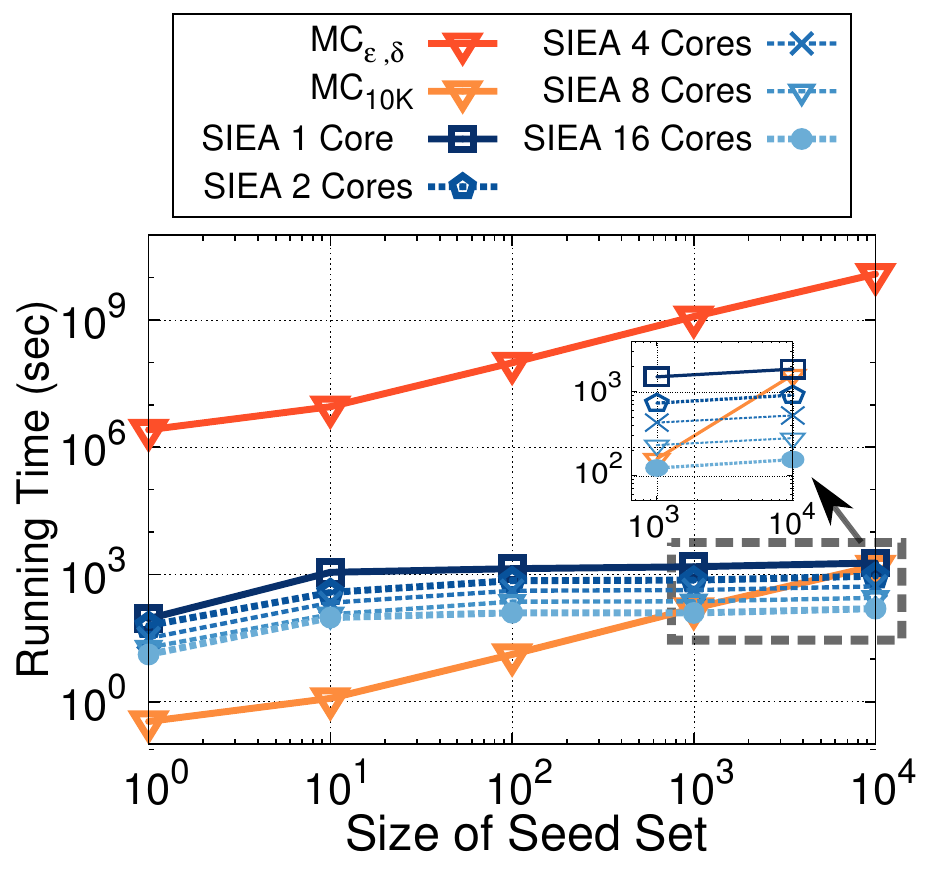}
	}
	\vspace{-0.1in}
	\caption{Comparing \OBInf{}, \MCk{} and \MCe{} on Twitter}
	\label{fig:set2c}
	\vspace{-0.2in}
\end{figure}

\setlength\tabcolsep{6pt}
\begin{table*}[!t] \small
	\centering
	\caption{ Comparing performance of algorithms in estimating influence spread in LT model (seed set size $|S|=1$)}   
	\vspace{-0.15in}
	\begin{tabular}{lrrrrrrrrrrrrr}
		\toprule
		\multirow{2}{*}{} & \multicolumn{3}{c}{\textbf{Avg. Rel. Error (\%)}} && \multicolumn{3}{c}{\textbf{Max. Rel. Error (\%)}} && \multicolumn{5}{c}{\textbf{Running time (sec.)}}  \\
		\cmidrule{2-4} \cmidrule{6-8} \cmidrule{10-14}
		\textbf{Dataset} & $\OBInf{}_{LT}$ & \MCk{} & \MCb{} && $\OBInf{}_{LT}$ & \MCk{} & \MCb{} && $\OBInf{}_{LT}$ & $\OBInf{}_{LT}$ \textsf{(16 cores)} & \MCk{} & \MCb{} & \MCe{} \\		
		\midrule	
		
		NetHEP & 1.6 & 1.6 & 0.6 && 8.4 & 7.9 & 2.5 && 0.0 & 0.0 & 0.0 & 0.1 & 1.0 \\ 
		NetPHY & 1.2 & 0.5 & 0.3 && 12.7 & 4.4 & 1.4 && 0.0 &  0.0 & 0.0 & 0.1 & 2.9 \\ 
		Epinions & 1.5 & 4.3 & 2.2 && 7.0 & 17.4 & 7.4 && 0.7 & 0.4 & 0.0 & 0.4 & 24.5 \\ 
		DBLP & 0.4 & 1.0 & 0.5 && 5.7 & 11.4 & 2.2 && 2.4 & 0.4 & 0.3 & 2.5 & 1530.4 \\ 
		Orkut & 0.5 & 3.3 & 1.1 && 1.9 & 22.1 & 5.9 && 249.4 & 25.0 & 8.5 & 84.2 & $4.6\cdot 10^4$ \\ 
		Twitter & 2.4 & 36.1 & 20.7 && 7.1 & 97.5 & 85.6 && 6820.0 & 548.6 & 32.2 & 287.6 & $1.4\cdot 10^7$ \\ 
		Friendster & 0.2 & 3.1 & 1.4 && 2.4 & 16.5 & 9.0 && 6183.9 & 701.8 & 20.4 & 137.8  & $9.3\cdot 10^6$ \\ 		
		\bottomrule		
	\end{tabular}
	\label{tab:infest_lt}
	\vspace{-0.1in}
\end{table*}

\vspace{-0.1in}
\subsection{Scalability Test}

We test the scalability of the single core and parallel versions of our method on synthetic networks generated by the well-known GTgraph with various network sizes. We also carry the same tests on the real-world Twitter network in comparison with the \MC{}.

\subsubsection{On Synthetic Datasets}

We generate synthetic graphs using GTgraph\cite{Bader06}, a standard graph generator used widely in large scale experiments on graph algorithms \cite{Hong11, Agarwal10, Campan08}. We generate graphs with number of nodes $n \in \{10^5, 10^6, 10^7, 10^8\}$. For each size $n$, we generate 3 different graphs with average degree $d \in \{10, 20, 30\}$. We use the \textsf{WC model} to assign edge weights. We run \OBInf{} with different number of cores $C=\{1,4,16\}$

Figure~\ref{fig:set2_syn} reports the time \OBInf{} spent to estimate influence spread of seed set of size 1. With the same number of nodes, we see that the running time of \OBInf{} does not significantly increase as the average degree increases.
Figure \ref{fig:set2_syn}b views Figure~\ref{fig:set2_syn}a in logarithmic scale to show the linear increase of running time with respect to the increases of nodes. As expected, \OBInf{} speeds up proportionally to number of cores used. As a result, \OBInf{} with 16 cores is able to estimate influence spread of a random node on a synthetic graph of 100 million nodes and 1.5 billion of edges in just 5 minutes.

\subsubsection{On Twitter Dataset}

Figure~\ref{fig:set2c} evaluates the performance of \OBInf{} in comparison with \MCk{} on various seed set sizes $|S|=\{1, 10, 100, 1k, 10k\}$ on Twitter dataset. On all the sizes of seed sets, \OBInf{} consistently has average and maximum relative errors smaller than 10\% (Figure \ref{fig:set2c}a). The maximum relative error of \MCk{} goes up to 244\% with seed set size $|S|=1$. As observed in experiments with large size seed sets, both \OBInf{} and \MCk{} have similar error rate with seed set size $|S|=10000$. 

In terms of running time, as the seed set size increases in powers of ten, \OBInf{}'s running time increases in much lower pace, e.g. few hundreds of seconds, while \MCe{} consumes proportionally more time (Figure~\ref{fig:set2c}b). Figure~\ref{fig:set2c}b also evaluates parallel implementation of \OBInf{} by varying number of CPU cores $C=\{1, 2, 4, 8, 16\}$. The running time of \OBInf{} reduces almost two times every time the number of cores doubles confirming the almost linear speedup.

Altogether, the parallel implementation of \OBInf{} shows a linear speedup behavior with respect to the number of cores used. On the same network with size of seed sets linearly grows, \OBInf{} requires slightly more time to estimate influence spread while Monte-Carlo shows a linear runtime requirement. Throughout the experiments, \OBInf{} always guarantees small error rate within $\epsilon$.

\vspace{-0.1in}
\subsection{Influence Estimation under LT Model}
We illustrate the generality of our algorithms in various diffusion model by adapting \OBInf{} for the LT model by only replacing \IICP{} with the sampling algorithm for the LT \cite{Kempe03}. The algorithm is then named $\OBInf{}_{LT}$. The setting is similar to the case of IC. We present the results of $\OBInf{}_{LT}$ compared with \MCk{}, \MCb{}, \MCe{} in Table~\ref{tab:infest_lt}. \textsf{INFEST} is initially proposed for the IC model, thus, we results for \textsf{INFEST} under the LT model are not available.

The results are mostly consistent with those observed under the IC model. $\OBInf{}_{LT}$ obtains significantly smaller errors and runs in order of magnitudes faster than the counterparts. The results again confirm that the estimation quality of \MC{} using $10K$ samples is not good enough to be considered as gold-standard quality benchmark.

%% file: body/conclusion.tex
\vspace{-0.15in}
\section{Conclusion}
\label{sec:con}
This paper investigates a new measure, called Outward Influence, for nodes' influence in social networks. Outward influence inspires new statiscal algorithms, namely Importance IC Polling (\IICP{}) and Robust Mean Estimation (\SiEIE) to estimate influence of nodes under various stochastic diffusion models. Under the popular IC model, the \IICP{} leads to an \FPRAS{} for estimating outward influence and \OBInf{} to estimate influence spread. \OBInf{} is $\Omega(\log^4(n))$ times faster than the most recent state-of-the-art and experimentally outperform the other methods by several orders of magnitudes. As previous approaches to compute ground truth influence can result in high error and long computational time, our algorithms provides concrete and scalable tools to estimate ground-truth influence for research on network cascade and social influence.

%% file: outward.bbl

\begin{thebibliography}{00}


\ifx \showCODEN    \undefined \def \showCODEN     #1{\unskip}     \fi
\ifx \showDOI      \undefined \def \showDOI       #1{{\tt DOI:}\penalty0{#1}\ }
  \fi
\ifx \showISBNx    \undefined \def \showISBNx     #1{\unskip}     \fi
\ifx \showISBNxiii \undefined \def \showISBNxiii  #1{\unskip}     \fi
\ifx \showISSN     \undefined \def \showISSN      #1{\unskip}     \fi
\ifx \showLCCN     \undefined \def \showLCCN      #1{\unskip}     \fi
\ifx \shownote     \undefined \def \shownote      #1{#1}          \fi
\ifx \showarticletitle \undefined \def \showarticletitle #1{#1}   \fi
\ifx \showURL      \undefined \def \showURL       #1{#1}          \fi
\providecommand\bibfield[2]{#2}
\providecommand\bibinfo[2]{#2}
\providecommand\natexlab[1]{#1}
\providecommand\showeprint[2][]{arXiv:#2}

\bibitem[\protect\citeauthoryear{Agarwal, Petrini, Pasetto, and Bader}{Agarwal
  et~al\mbox{.}}{2010}]%
        {Agarwal10}
\bibfield{author}{\bibinfo{person}{V. Agarwal}, \bibinfo{person}{F. Petrini},
  \bibinfo{person}{D. Pasetto}, {and} \bibinfo{person}{D.~A. Bader}.}
  \bibinfo{year}{2010}\natexlab{}.
\newblock \showarticletitle{Scalable graph exploration on multicore
  processors}. In \bibinfo{booktitle}{{\em SC}}. IEEE, \bibinfo{pages}{1--11}.
\newblock


\bibitem[\protect\citeauthoryear{Bader and Madduri}{Bader and Madduri}{2006}]%
        {Bader06}
\bibfield{author}{\bibinfo{person}{D.~A. Bader} {and} \bibinfo{person}{K.
  Madduri}.} \bibinfo{year}{2006}\natexlab{}.
\newblock \showarticletitle{Gtgraph: A synthetic graph generator suite}.
\newblock \bibinfo{journal}{{\em Atlanta, GA, February\/}}
  (\bibinfo{year}{2006}).
\newblock


\bibitem[\protect\citeauthoryear{Borgs, Brautbar, Chayes, and Lucier}{Borgs
  et~al\mbox{.}}{2014}]%
        {Borgs14}
\bibfield{author}{\bibinfo{person}{C. Borgs}, \bibinfo{person}{M. Brautbar},
  \bibinfo{person}{J. Chayes}, {and} \bibinfo{person}{B. Lucier}.}
  \bibinfo{year}{2014}\natexlab{}.
\newblock \showarticletitle{Maximizing Social Influence in Nearly Optimal
  Time}. In \bibinfo{booktitle}{{\em SODA}}. \bibinfo{publisher}{SIAM},
  \bibinfo{pages}{946--957}.
\newblock


\bibitem[\protect\citeauthoryear{Campan}{Campan}{2008}]%
        {Campan08}
\bibfield{author}{\bibinfo{person}{A. Campan}.}
  \bibinfo{year}{2008}\natexlab{}.
\newblock \showarticletitle{A clustering approach for data and structural
  anonymity in social networks}.
\newblock \bibinfo{journal}{{\em PinKDD\/}} (\bibinfo{year}{2008}),
  \bibinfo{pages}{54}.
\newblock


\bibitem[\protect\citeauthoryear{Cha, Mislove, and Gummadi}{Cha
  et~al\mbox{.}}{2009}]%
        {Cha09}
\bibfield{author}{\bibinfo{person}{M. Cha}, \bibinfo{person}{A. Mislove}, {and}
  \bibinfo{person}{K.~P. Gummadi}.} \bibinfo{year}{2009}\natexlab{}.
\newblock \showarticletitle{A measurement-driven analysis of information
  propagation in the flickr social network}. In \bibinfo{booktitle}{{\em WWW}}.
  \bibinfo{publisher}{ACM}, \bibinfo{pages}{721--730}.
\newblock


\bibitem[\protect\citeauthoryear{Chen, Wang, and Wang}{Chen
  et~al\mbox{.}}{2010}]%
        {Chen10}
\bibfield{author}{\bibinfo{person}{W. Chen}, \bibinfo{person}{C. Wang}, {and}
  \bibinfo{person}{Y. Wang}.} \bibinfo{year}{2010}\natexlab{}.
\newblock \showarticletitle{Scalable influence maximization for prevalent viral
  marketing in large-scale social networks}. In \bibinfo{booktitle}{{\em KDD}}.
  \bibinfo{publisher}{ACM}, \bibinfo{pages}{1029--1038}.
\newblock


\bibitem[\protect\citeauthoryear{Chung and Lu}{Chung and Lu}{2006}]%
        {Chung06}
\bibfield{author}{\bibinfo{person}{F. Chung} {and} \bibinfo{person}{L. Lu}.}
  \bibinfo{year}{2006}\natexlab{}.
\newblock \showarticletitle{Concentration inequalities and martingale
  inequalities: a survey}.
\newblock \bibinfo{journal}{{\em Internet Mathematics\/}}
  (\bibinfo{year}{2006}), \bibinfo{pages}{79--127}.
\newblock


\bibitem[\protect\citeauthoryear{Cohen, Delling, Pajor, and Werneck}{Cohen
  et~al\mbox{.}}{2014}]%
        {Cohen14}
\bibfield{author}{\bibinfo{person}{E. Cohen}, \bibinfo{person}{D. Delling},
  \bibinfo{person}{T. Pajor}, {and} \bibinfo{person}{R.~F. Werneck}.}
  \bibinfo{year}{2014}\natexlab{}.
\newblock \showarticletitle{Sketch-based influence maximization and
  computation: Scaling up with guarantees}. In \bibinfo{booktitle}{{\em CIKM}}.
  ACM, \bibinfo{pages}{629--638}.
\newblock


\bibitem[\protect\citeauthoryear{Dagum, Karp, Luby, and Ross}{Dagum
  et~al\mbox{.}}{2000}]%
        {Dagum00}
\bibfield{author}{\bibinfo{person}{P. Dagum}, \bibinfo{person}{R. Karp},
  \bibinfo{person}{M. Luby}, {and} \bibinfo{person}{S. Ross}.}
  \bibinfo{year}{2000}\natexlab{}.
\newblock \showarticletitle{An Optimal Algorithm for Monte Carlo Estimation}.
\newblock \bibinfo{journal}{{\em SICOMP\/}} (\bibinfo{year}{2000}),
  \bibinfo{pages}{1484--1496}.
\newblock
\showISSN{0097-5397}


\bibitem[\protect\citeauthoryear{Daley, Gani, and Gani}{Daley
  et~al\mbox{.}}{2001}]%
        {Daley01}
\bibfield{author}{\bibinfo{person}{D.~J. Daley}, \bibinfo{person}{J. Gani},
  {and} \bibinfo{person}{J.~M. Gani}.} \bibinfo{year}{2001}\natexlab{}.
\newblock \bibinfo{booktitle}{{\em Epidemic modelling: an introduction}}.
  Vol.~\bibinfo{volume}{15}.
\newblock \bibinfo{publisher}{Cambridge University Press}.
\newblock


\bibitem[\protect\citeauthoryear{Dinh and Thai}{Dinh and Thai}{2015}]%
        {Dinh15infocom}
\bibfield{author}{\bibinfo{person}{T.~N. Dinh} {and} \bibinfo{person}{M.~T.
  Thai}.} \bibinfo{year}{2015}\natexlab{}.
\newblock \showarticletitle{Assessing attack vulnerability in networks with
  uncertainty}. In \bibinfo{booktitle}{{\em INFOCOM}}. IEEE,
  \bibinfo{pages}{2380--2388}.
\newblock


\bibitem[\protect\citeauthoryear{Du, Song, Gomez-Rodriguez, and Zha}{Du
  et~al\mbox{.}}{2013}]%
        {Du13}
\bibfield{author}{\bibinfo{person}{N. Du}, \bibinfo{person}{L. Song},
  \bibinfo{person}{M. Gomez-Rodriguez}, {and} \bibinfo{person}{H. Zha}.}
  \bibinfo{year}{2013}\natexlab{}.
\newblock \showarticletitle{Scalable influence estimation in continuous-time
  diffusion networks}. In \bibinfo{booktitle}{{\em NIPS}}.
  \bibinfo{pages}{3147--3155}.
\newblock


\bibitem[\protect\citeauthoryear{Goyal, Bonchi, and Lakshmanan}{Goyal
  et~al\mbox{.}}{2010}]%
        {Goyal10}
\bibfield{author}{\bibinfo{person}{A. Goyal}, \bibinfo{person}{F. Bonchi},
  {and} \bibinfo{person}{L.~V.~S. Lakshmanan}.}
  \bibinfo{year}{2010}\natexlab{}.
\newblock \showarticletitle{Learning Influence Probabilities in Social
  Networks}. In \bibinfo{booktitle}{{\em WSDM}}. \bibinfo{publisher}{ACM},
  \bibinfo{pages}{241--250}.
\newblock


\bibitem[\protect\citeauthoryear{Gupta, Goel, Lin, Sharma, Wang, and
  Zadeh}{Gupta et~al\mbox{.}}{2013}]%
        {Gupta13}
\bibfield{author}{\bibinfo{person}{P. Gupta}, \bibinfo{person}{A. Goel},
  \bibinfo{person}{J. Lin}, \bibinfo{person}{A. Sharma}, \bibinfo{person}{D.
  Wang}, {and} \bibinfo{person}{R. Zadeh}.} \bibinfo{year}{2013}\natexlab{}.
\newblock \showarticletitle{Wtf: The who to follow service at twitter}. In
  \bibinfo{booktitle}{{\em WWW}}. ACM, \bibinfo{pages}{505--514}.
\newblock


\bibitem[\protect\citeauthoryear{Hong, Kim, Oguntebi, and Olukotun}{Hong
  et~al\mbox{.}}{2011}]%
        {Hong11}
\bibfield{author}{\bibinfo{person}{S. Hong}, \bibinfo{person}{Sang~K. Kim},
  \bibinfo{person}{T. Oguntebi}, {and} \bibinfo{person}{K. Olukotun}.}
  \bibinfo{year}{2011}\natexlab{}.
\newblock \showarticletitle{Accelerating CUDA graph algorithms at maximum
  warp}. In \bibinfo{booktitle}{{\em SIGPLAN Notices}}. ACM,
  \bibinfo{pages}{267--276}.
\newblock


\bibitem[\protect\citeauthoryear{Ioannides and Datcher}{Ioannides and
  Datcher}{2004}]%
        {Ioannides04}
\bibfield{author}{\bibinfo{person}{Y.~M. Ioannides} {and}
  \bibinfo{person}{L.~L. Datcher}.} \bibinfo{year}{2004}\natexlab{}.
\newblock \showarticletitle{Job information networks, neighborhood effects, and
  inequality}.
\newblock \bibinfo{journal}{{\em Journal of economic literature\/}}
  (\bibinfo{year}{2004}), \bibinfo{pages}{1056--1093}.
\newblock


\bibitem[\protect\citeauthoryear{Kempe, Kleinberg, and Tardos}{Kempe
  et~al\mbox{.}}{2003}]%
        {Kempe03}
\bibfield{author}{\bibinfo{person}{D. Kempe}, \bibinfo{person}{J. Kleinberg},
  {and} \bibinfo{person}{{\'E}. Tardos}.} \bibinfo{year}{2003}\natexlab{}.
\newblock \showarticletitle{Maximizing the spread of influence through a social
  network}. In \bibinfo{booktitle}{{\em KDD}}. \bibinfo{pages}{137--146}.
\newblock


\bibitem[\protect\citeauthoryear{Kempe, Kleinberg, and Tardos}{Kempe
  et~al\mbox{.}}{2005}]%
        {Kempe05}
\bibfield{author}{\bibinfo{person}{D. Kempe}, \bibinfo{person}{J. Kleinberg},
  {and} \bibinfo{person}{E. Tardos}.} \bibinfo{year}{2005}\natexlab{}.
\newblock \showarticletitle{Influential nodes in a diffusion model for social
  networks}. In \bibinfo{booktitle}{{\em ICALP}}. \bibinfo{pages}{1127--1138}.
\newblock


\bibitem[\protect\citeauthoryear{Krause, Singh, and Guestrin}{Krause
  et~al\mbox{.}}{2008}]%
        {Krause08}
\bibfield{author}{\bibinfo{person}{A. Krause}, \bibinfo{person}{A. Singh},
  {and} \bibinfo{person}{C. Guestrin}.} \bibinfo{year}{2008}\natexlab{}.
\newblock \showarticletitle{Near-optimal sensor placements in Gaussian
  processes: Theory, efficient algorithms and empirical studies}.
\newblock \bibinfo{journal}{{\em JMLR\/}} (\bibinfo{year}{2008}),
  \bibinfo{pages}{235--284}.
\newblock


\bibitem[\protect\citeauthoryear{Kwak, Lee, Park, and Moon}{Kwak
  et~al\mbox{.}}{2010}]%
        {Kwak10}
\bibfield{author}{\bibinfo{person}{H. Kwak}, \bibinfo{person}{C. Lee},
  \bibinfo{person}{H. Park}, {and} \bibinfo{person}{S. Moon}.}
  \bibinfo{year}{2010}\natexlab{}.
\newblock \showarticletitle{What is Twitter, a social network or a news
  media?}. In \bibinfo{booktitle}{{\em WWW}}. ACM, \bibinfo{pages}{591--600}.
\newblock


\bibitem[\protect\citeauthoryear{Leskovec, Krause, Guestrin, Faloutsos,
  VanBriesen, and Glance}{Leskovec et~al\mbox{.}}{2007}]%
        {Leskovec07}
\bibfield{author}{\bibinfo{person}{J. Leskovec}, \bibinfo{person}{A. Krause},
  \bibinfo{person}{C. Guestrin}, \bibinfo{person}{C. Faloutsos},
  \bibinfo{person}{J. VanBriesen}, {and} \bibinfo{person}{N. Glance}.}
  \bibinfo{year}{2007}\natexlab{}.
\newblock \showarticletitle{Cost-effective outbreak detection in networks}. In
  \bibinfo{booktitle}{{\em KDD}}. \bibinfo{publisher}{ACM},
  \bibinfo{pages}{420--429}.
\newblock


\bibitem[\protect\citeauthoryear{Lin and Schatz}{Lin and Schatz}{2010}]%
        {Lin10}
\bibfield{author}{\bibinfo{person}{J. Lin} {and} \bibinfo{person}{M. Schatz}.}
  \bibinfo{year}{2010}\natexlab{}.
\newblock \showarticletitle{Design patterns for efficient graph algorithms in
  MapReduce}. In \bibinfo{booktitle}{{\em MLG}}. ACM, \bibinfo{pages}{78--85}.
\newblock


\bibitem[\protect\citeauthoryear{Lucier, Oren, and Singer}{Lucier
  et~al\mbox{.}}{2015}]%
        {Lucier15}
\bibfield{author}{\bibinfo{person}{B. Lucier}, \bibinfo{person}{J. Oren}, {and}
  \bibinfo{person}{Y. Singer}.} \bibinfo{year}{2015}\natexlab{}.
\newblock \showarticletitle{Influence at scale: Distributed computation of
  complex contagion in networks}. In \bibinfo{booktitle}{{\em KDD}}. ACM,
  \bibinfo{pages}{735--744}.
\newblock


\bibitem[\protect\citeauthoryear{Mitzenmacher and Upfal}{Mitzenmacher and
  Upfal}{2005}]%
        {Mitzenmacher05}
\bibfield{author}{\bibinfo{person}{M. Mitzenmacher} {and} \bibinfo{person}{E.
  Upfal}.} \bibinfo{year}{2005}\natexlab{}.
\newblock \bibinfo{booktitle}{{\em Probability and computing: Randomized
  algorithms and probabilistic analysis}}.
\newblock \bibinfo{publisher}{Cambridge University Press}.
\newblock


\bibitem[\protect\citeauthoryear{Nguyen, Huiyuan, Das, Thai, and Dinh}{Nguyen
  et~al\mbox{.}}{2013}]%
        {Nguyen13icdm}
\bibfield{author}{\bibinfo{person}{D.~T. Nguyen}, \bibinfo{person}{Z. Huiyuan},
  \bibinfo{person}{S. Das}, \bibinfo{person}{M.~T. Thai}, {and}
  \bibinfo{person}{T.~N. Dinh}.} \bibinfo{year}{2013}\natexlab{}.
\newblock \showarticletitle{Least Cost Influence in Multiplex Social Networks:
  Model Representation and Analysis}. In \bibinfo{booktitle}{{\em ICDM}}.
  \bibinfo{pages}{567--576}.
\newblock


\bibitem[\protect\citeauthoryear{Nguyen, Ghosh, Mayo, and Dinh}{Nguyen
  et~al\mbox{.}}{2016}]%
        {Nguyen165}
\bibfield{author}{\bibinfo{person}{H.~T. Nguyen}, \bibinfo{person}{P. Ghosh},
  \bibinfo{person}{M.~L. Mayo}, {and} \bibinfo{person}{T.~N. Dinh}.}
  \bibinfo{year}{2016}\natexlab{}.
\newblock \showarticletitle{Multiple Infection Sources Identification with
  Provable Guarantees}. In \bibinfo{booktitle}{{\em CIKM}}. ACM,
  \bibinfo{pages}{1663--1672}.
\newblock


\bibitem[\protect\citeauthoryear{Nguyen, Thai, and Dinh}{Nguyen
  et~al\mbox{.}}{2016a}]%
        {Nguyen162}
\bibfield{author}{\bibinfo{person}{H.~T. Nguyen}, \bibinfo{person}{M.~T. Thai},
  {and} \bibinfo{person}{T.~N. Dinh}.} \bibinfo{year}{2016}\natexlab{a}.
\newblock \showarticletitle{Cost-aware targeted viral marketing in
  billion-scale networks}. In \bibinfo{booktitle}{{\em INFOCOM}}. IEEE,
  \bibinfo{pages}{1--9}.
\newblock


\bibitem[\protect\citeauthoryear{Nguyen, Thai, and Dinh}{Nguyen
  et~al\mbox{.}}{2016b}]%
        {Nguyen163}
\bibfield{author}{\bibinfo{person}{H.~T. Nguyen}, \bibinfo{person}{M.~T. Thai},
  {and} \bibinfo{person}{T.~N. Dinh}.} \bibinfo{year}{2016}\natexlab{b}.
\newblock \showarticletitle{Stop-and-Stare: Optimal Sampling Algorithms for
  Viral Marketing in Billion-scale Networks}. In \bibinfo{booktitle}{{\em
  SIGMOD}}. \bibinfo{publisher}{ACM}, \bibinfo{pages}{695--710}.
\newblock


\bibitem[\protect\citeauthoryear{Ohsaka, Akiba, Yoshida, and
  Kawarabayashi}{Ohsaka et~al\mbox{.}}{2016}]%
        {Ohsaka16}
\bibfield{author}{\bibinfo{person}{N. Ohsaka}, \bibinfo{person}{T. Akiba},
  \bibinfo{person}{Y. Yoshida}, {and} \bibinfo{person}{K. Kawarabayashi}.}
  \bibinfo{year}{2016}\natexlab{}.
\newblock \showarticletitle{Dynamic influence analysis in evolving networks}.
\newblock \bibinfo{journal}{{\em VLDB\/}} (\bibinfo{year}{2016}),
  \bibinfo{pages}{1077--1088}.
\newblock


\bibitem[\protect\citeauthoryear{Preciado, Zargham, Enyioha, Jadbabaie, and
  Pappas}{Preciado et~al\mbox{.}}{2013}]%
        {Preciado13}
\bibfield{author}{\bibinfo{person}{V.~M. Preciado}, \bibinfo{person}{M.
  Zargham}, \bibinfo{person}{C. Enyioha}, \bibinfo{person}{A. Jadbabaie}, {and}
  \bibinfo{person}{G. Pappas}.} \bibinfo{year}{2013}\natexlab{}.
\newblock \showarticletitle{Optimal vaccine allocation to control epidemic
  outbreaks in arbitrary networks}. In \bibinfo{booktitle}{{\em CDC}}. IEEE,
  \bibinfo{pages}{7486--7491}.
\newblock


\bibitem[\protect\citeauthoryear{Tang, Shi, and Xiao}{Tang
  et~al\mbox{.}}{2015}]%
        {Tang15}
\bibfield{author}{\bibinfo{person}{Y. Tang}, \bibinfo{person}{Y. Shi}, {and}
  \bibinfo{person}{X. Xiao}.} \bibinfo{year}{2015}\natexlab{}.
\newblock \showarticletitle{Influence Maximization in Near-Linear Time: A
  Martingale Approach}. In \bibinfo{booktitle}{{\em SIGMOD}}. ACM,
  \bibinfo{pages}{1539--1554}.
\newblock


\bibitem[\protect\citeauthoryear{Tang, Xiao, and Shi}{Tang
  et~al\mbox{.}}{2014}]%
        {Tang14}
\bibfield{author}{\bibinfo{person}{Y. Tang}, \bibinfo{person}{X. Xiao}, {and}
  \bibinfo{person}{Y. Shi}.} \bibinfo{year}{2014}\natexlab{}.
\newblock \showarticletitle{Influence maximization: Near-optimal time
  complexity meets practical efficiency}. In \bibinfo{booktitle}{{\em SIGMOD}}.
  ACM, \bibinfo{pages}{75--86}.
\newblock


\end{thebibliography}
